\documentclass[a4paper,11pt]{article}
\usepackage{aaskaiid}
\usepackage{orcidlink}

\usepackage{xspace}
\usepackage{gensymb}

\title{Dual AGN and Multiple SMBH Systems in the Era of SKAO}
\ShortTitle{Dual AGN and Multiple SMBH Systems in the Era of SKAO}

\author[1\ast]{Quirino D'Amato\orcidlink{0000-0002-9948-0897}}
\ShortName{D'Amato et al.}
\author[2]{Lang Cui\orcidlink{0000-0003-0721-5509}}
\author[3]{Roger Deane}
\author[4]{S. Komossa}
\author[3]{Coral Pillay}
\author[2]{Ashutosh Tripathi}
\author[5]{Preeti Kharb}
\author[6]{Hengxiao Guo}
\author[7]{Sumana Nandi}
\author[8]{Khatun Rubinur\orcidlink{0000-0001-5574-5104}}
\author[9]{Sonia Anton}
\author[6]{Tao An \orcidlink{0000-0003-4341-0029}}
\author[10]{Silvia Bonoli}
\author[2]{Ning Chang}
\author[11]{Romeel Dave}
\author[12]{Alessandra De Rosa}
\author[13]{Melanie Habouzit}
\author[1]{Filippo Mannucci}
\author[14]{Isabella Prandoni}
\author[15]{Paola Severgnini}
\author[16]{Martina Scialpi}
\author[14]{Cristiana Spingola}
\author[17]{Cristian Vignali}
\author[2]{Wancheng Xu}
\author[2]{Xi Yan}
\author[6]{Yingkang Zhang}

\affiliation[1]{INAF – Osservatorio Astrofisico di Arcetri, Via Largo E. Fermi 5, Firenze 50125, Italy}
\affiliation[2]{Xinjiang Astronomical Observatory, CAS, 150 Science-1 Street, Urumqi 830011, China}
\affiliation[3]{Wits Centre for Astrophysics, University of the Witwatersrand, 1 Jan Smuts Avenue, Johannesburg, 2000, South Africa}
\affiliation[4]{Max-Planck-Institut für Radioastronomie, Auf dem Hügel 69, 53121 Bonn, Germany}
\affiliation[5]{National Centre for Radio Astrophysics (NCRA) - Tata Institute of Fundamental Research (TIFR), S. P. Pune University Campus, Ganeshkhind, Pune 411007, India}
\affiliation[6]{Shanghai Astronomical Observatory, CAS, 80 Nandan Road, Shanghai, China}
\affiliation[7]{Institute of Astronomy Space and Earth Sciences, P 177, CIT Road, Scheme 7m, Kolkata 700054, West Bengal, India }
\affiliation[8]{Institute of Theoretical Astrophysics, University of Oslo, P.O. Box 1029, Blindern, 0315 Oslo, Norway}
\affiliation[9]{CFisUC, Departamento de Física, Universidade de Coimbra, 3004-516 Coimbra, Portugal}
\affiliation[10]{Donostia International Physics Center (DIPC), Paseo Manuel de Lardizabal 4, 20018 Donostia-San Sebastian, Spain \& IKERBASQUE, Basque Foundation for Science, E-48013, Bilbao, Spain}
\affiliation[11]{Institute for Astronomy, Royal Observatory, University of Edinburgh, Edinburgh EH9 3HJ, UK}
\affiliation[12]{INAF - Istituto di Astrofisica e Planetologia Spaziali, Via del Fosso del Cavaliere I-00133, Roma, Italy}
\affiliation[13]{Department of Astronomy, University of Geneva, Chemin d'Ecogia, CH-1290 Versoix, Switzerland}
\affiliation[14]{INAF - Istituto di Radioastronomia, Via P. Gobetti 101, I$-$40129, Italy}
\affiliation[15]{INAF - Osservatorio Astronomico di Brera, via Brera 28, I-20121 Milano, Italy \& via Bianchi 46, Merate (LC), Italy}
\affiliation[16]{Dipartimento di Fisica e Astronomia, Università di Firenze, via G. Sansone 1, 50019 Sesto F.no, Firenze, Italy}
\affiliation[17]{Dipartimento di Fisica e Astronomia, Università degli Studi di Bologna, Via P. Gobetti 93/2, 40129, Bologna, Italy}

\affiliation[\ast]{Corresponding author: Email: quirino.damato@inaf.it}

\abstract{We present a radio-oriented review of current strategies for the detection and characterization of dual active galactic nuclei (DAGN) and supermassive black holes binaries (SMBHBs), emphasizing the crucial role of radio interferometry in advancing this field. We discuss how high-resolution radio imaging -- particularly through very long baseline interferometry (VLBI) -- provides an unique, dust-unbiased tool to identify multiple accreting SMBHs, disentangle AGN-related emission from star formation, and trace components from tens of kpc to sub-parsec scales. We summarize current observational limitations, such as insufficient sensitivity–resolution combination and area coverage. 
We then outline how the SKAO will overcome these constraints through its unprecedented combination of sensitivity, survey speed, imaging fidelity and angular resolution, enabling the discovery and characterization of dual and binary SMBHs from the nearby Universe to the epoch of reionization. Several science cases are presented, including radio follow-ups of optical/infrared-selected DAGN, direct blind radio selection of DAGN, studies of compact bound SMBHBs, and the link between SMBHB orbital evolution and low-frequency gravitational wave emission. We further emphasize the synergy between SKAO observations and modern and upcoming facilities such as the James Webb and Euclid space telescopes, Rubin observatory, and gravitational wave detectors including the laser interferometer space antenna and pulsar timing arrays. These combined capabilities will allow SKAO to enable the first comprehensive radio census of dual and binary SMBH systems, bridge the gap between electromagnetic and gravitational wave observations, and provide a statistically significant view of SMBH pairing, accretion, and merger-driven feedback throughout cosmic history.}

\begin{document}
\maketitle

\section{Introduction}

Super-massive black holes (SMBHs), with masses of $\sim10^6 - 10^9 ~\mathrm{M_\odot}$, are believed to reside ubiquitously in galactic centers, particularly in the bulge of elliptical and spiral galaxies \citep[see][and references therein]{kormendy_1995, derosa_2019}. The tight observational correlations between SMBHs masses and properties of their host galaxies, such as stellar velocity dispersion, strongly suggest a fundamental co-evolution between SMBHs and their galactic environments over cosmic timescales \citep{KH_2013}. A primary mechanism driving both galaxy and SMBH growth is hierarchical merging: hydrodynamical simulations show that galaxy major mergers produce gas inflows towards the center of galaxies, which can trigger both star formation (SF) and accretion onto their central SMBHs, producing luminous active galactic nuclei \cite[AGN,][]{DiMatteo2005, mayer_2007}. As a result of the inspiraling merging process, multi-SMBH systems form \citep{begelman_1980,volonteri_2003}. In particular, dual AGN (DAGN) are systems where two active SMBHs have a separation larger than their radii of gravitational influence, but still evolving within the potential of the merged host galaxies \citep{merritt_2013,volonteri_2022}. These objects, typically found at hundreds of pc up to tens of kpc separation, are considered the direct precursors of SMBH binaries (SMBHBs), where two SMBHs are gravitationally bound, usually at pc to sub-pc scales \citep[$\lesssim$10 pc,][]{mayer_2007,dotti_2007}. 

The study of these multi-SMBH systems is of paramount importance for several reasons. They offer crucial insights into the intricate interplay between galaxy mergers and SMBHs fueling and growth, enabling tests of the many predictions of galaxy evolution models. Examples of testable predictions are their separation distribution, their number density, fraction in the overall AGN population, redshift evolution, the mass and luminosity ratios between the primary and secondary components, the impact on host galaxies, and many others \citep{volonteri_2022,shen_2023,dimatteo_2023}. Moreover, SMBHBs are predicted to be the primary sources of low-frequency (nHz-$\mu$Hz) gravitational waves (GWs) in the Universe \citep{jaffe_2003,sesana_2013, colpi_2019}, making them key targets for GW observatories such as the laser interferometer space antenna (LISA, \citealt{Amaro-Seoane_2023}) and pulsar timing arrays \citep[PTAs,][]{agazie_2023,EPTA_collab_2023}.

Observationally, however, directly detecting and characterizing these systems, especially at small separations and high redshifts, remains a significant challenge due to their inherent rarity, and the large sky area coverage and resolution required to select and confirm a significant amount of candidates (\citealt{derosa_2019}; see also Sec. \ref{sec:intro_DAGN} and Sec. \ref{sec:intro_BinBH}). Radio emission is an unique tool to identify multiple SMBHs candidates and confirm their nature; it is not affected by dust obscuration \citep{deane2014nature}, it can be observed down to sub-arcsecond resolution thanks to the interferometry technique (milli-arcsecond for very long baseline interferometry, VLBI) and it is not affected by spurious emission of gas photo-ionization region ``mimicking'' a second AGN like in optical data \citep{fu_2012}. However, current cm-wavelength surveys with sub-arcsec resolution are limited by either shallow sensitivity and/or small area coverage. The advent of next-generation facilities, particularly the Square Kilometre Array Observatory (SKAO), promises to revolutionize this field by providing an unparalleled ability to discover, characterize, and study multi-SMBH systems across a vast range of cosmic distance and orbital separation, thanks to the unmatched sensitivity, survey speed and imaging fidelity, combined with the high angular resolution that it will achieve.

\subsection{The quest of finding high-$z$ DAGN}
\label{sec:intro_DAGN}
\begin{figure}[ht]
    \centering
	\includegraphics[width=1\columnwidth]{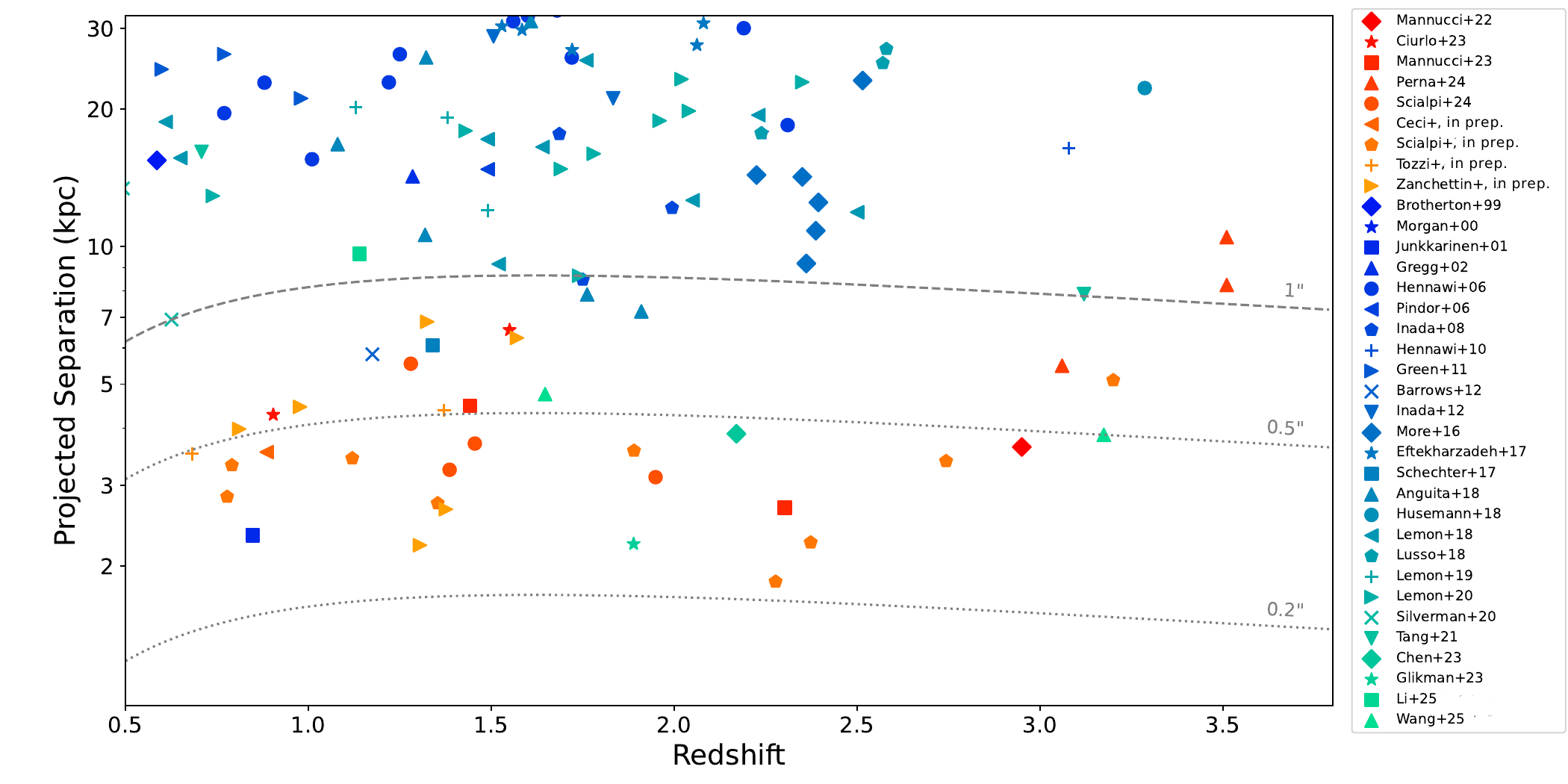}
    \caption{ Projected separations of all spectroscopically confirmed high-$z$ DAGN, identified at projected separation $\leq30$ kpc and in the $0.5\leq z \leq 3.5$ range. Updated from \cite{mannucci_2023}.}
    \label{fig:DAGN_census}
\end{figure}

Given the large galaxy merger timescale \citep[$\sim$1 Gyr,][]{tremmel_2017}, the presence of DAGN is expected for $\gtrsim$ 1\% of bright ($L_{\mathrm{bol}}>10^{43}~\mathrm{erg~s^{-1}}$) AGN at $z\gtrsim$2 \citep{volonteri_2003,volonteri_2022}. While many dual systems at $z>0.5$ are needed to robustly test model predictions, identifying these systems is a difficult task because it requires large sky area coverage with sub-arcsec resolution \citep{derosa_2019}. X-ray emission has been used in the past to select these systems, but due to low photon-counting statistics and coarse resolution in wide area surveys, the detection/characterization of candidates is practically possible only in the nearby Universe \citep{derosa_2019}. Currently, the most promising selection techniques rely on the optical emission of relatively bright quasars \citep[$G_\mathrm{mag} \lesssim$ 21;][]{shen_2019,mannucci_2022}. In the last few years, techniques based on the precise astrometry and the high resolution offered by the \textit{Gaia} all-sky survey \citep{gaia_collab_2024} emerged as efficient methods to select high-$z$ DAGN. In particular, the ``varstrometry'' and \textit{Gaia}-multi-peak (GMP) selection criteria allowed to discover tens of new dual systems at $z>0.5$ down to few kpc projected separation. The ``varstrometry'' technique selects DAGN candidates on the basis of the extra-astrometric noise produced by luminosity variability of the members of an otherwise unresolved dual system \citep{shen_2019}. The GMP method, instead, searches for multiple peaks in the light profiles of optically/infrared-selected
quasars in the \textit{Gaia} archive \citep{mannucci_2022}; this technique led to the selection of hundreds of DAGN candidates with sub-arcsecond separation, between 0.15'' and $\sim$0.7'' \citep[e.g.][]{mannucci_2023,ciurlo_2023,scialpi_2024}. A complete census of confirmed DAGN at $z>0.5$ and projected separation $<$30 kpc is shown in Fig. \ref{fig:DAGN_census}.

Despite the promises offered by these novel techniques, time-expensive spectroscopy of each target is still required to confirm multiple SMBHs systems, since optically selected objects could be in general associated to DAGN, gravitationally-lensed systems (GLS), or a projected alignment between an AGN and a foreground star. In this context, follow-up observations in the radio band have increasingly become a standard method for uncovering the nature of optically selected systems. Radio compact cores detected in correspondence of the optical/infrared centroids are an effective method to strengthen the DAGN nature of the candidates \citep{glikman_2023}. 
Radio observations are also useful for excluding GLS contaminants by comparing the optical morphology and multi-band flux ratios of the components (see Sec. \ref{sec:case_1}). In addition, cross-matching \textit{Gaia} and radio VLBI astrometry has proven beneficial for studying quasar jet systems \citep[e.g.][]{makarov_2017,plavin_2019,petrov_2019} and for identifying binary AGN \citep{breiding_2021}. Candidate binary AGN can also be identified via radio-optical positional offsets \citep{skipper_browne_2018}.

The selection of DAGN candidates directly in the radio band has the potential to add hundreds of new DAGN candidates, and to uncover the hidden (and mostly unknown) population of obscured DAGN \citep{deane2014nature}. Modern interferometers can reach $\sim\mu$Jy/beam sensitivity and cover large sky areas with modest time investment with respect to other wavelengths. Radio AGN are traditionally classified as radio-loud (RL) or radio-quiet (RQ) based on the ratio between their radio and optical (or X-ray) emission \citep{terashima_2003}; the RQ AGN usually overcome the RL population at $\lesssim 100\ \mu$Jy \citep{bonzini_2013}. Thanks to the sensitivity of the new radio surveys, also this faint RQ population, traditionally missed by previous generation, shallower wide-field surveys, has started to be detected and studied \citep{damato_2022,hale_2025}.
Furthermore, the James Webb space telescope (JWST) is currently expanding the high-$z$ frontier, including important discoveries of high-$z$ AGN and highest-redshift galaxy pairs and mergers \citep{Duan2025,perna_2025}. As many of the mergers in the high redshift universe are highly obscured, sensitive radio observations provide an additional or alternative route of finding AGN in these systems. However, the origin of radio emission is unclear: when the $\sim 100~ \mu$Jy regime is approached, it could be ascribed to both AGN or SF activity. In this regime, the radio-excess criterion (i.e., the radio emission in excess with respect to that expected from SF) is commonly adopted to select AGN, yielding a fraction of radio-excess sources of approximately 30\% \citep{smolcic_2017}.

While VLBI provides the milli-arcsecond resolution necessary to resolve compact DAGN, it remains technically constrained for broad demographic studies. Current VLBI observations are primarily limited to bright sources ($>$1 mJy at 1.4 GHz), and up to 70\% of the radio flux can be resolved out when larger-scale structures, such as jets, are present \citep{spingola_2020}. Furthermore, the narrow, arcsecond-scale field of view (FOV) of current VLBI arrays makes them unsuitable for blind surveys. As a result, the radio properties of faint, high-$z$ DAGN have not yet been explored in statistically significant, homogeneous samples \citep[e.g.,][]{fu_2012, deane2014nature,spingola2019,glikman_2023,xu_2024,schwartzman_2024}. Nevertheless, in some low-redshift DAGN with sub-kpc separations, VLBI remains an effective tool for identifying multiple BHs (see Sec. \ref{sec:intro_BinBH}).

An effective way to distinguish between AGN and SF radio emission is to measure the equivalent brightness temperature $T_b$ \citep{condon_1980}. AGN typically exhibit $T_b \gtrsim 10^5$ K at 1.4 GHz, with the threshold primarily set by observing frequency \citep{condon_1980,morabito_2022}. While milli-arcsecond resolution is required at GHz frequencies, the higher threshold at 144 MHz ($T_b \sim 10^6$ K) allows AGN identification at sub-arcsecond resolution scales. This method has enabled the efficient selection of hundreds of radio AGN in LOFAR/LoTSS using international baselines \citep{shimwell_2017,morabito_2022}.

\subsection{Local DAGN and compact bound binary SMBHs}
\label{sec:intro_BinBH}

\subsubsection{Local DAGN at the centers of single galaxies, still spatially resolved}

A number of wide-separation SMBH pairs located at the centers of {\em single} galaxies or very advanced mergers have been identified in the nearby universe using hard X-ray or radio observations \citep[e.g.,][]{komossa_2003, deane2014nature, Rubinur2019}. These systems are challenging to identify at optical wavelengths because the vast majority of gas-rich mergers are heavily obscured by large columns of dusty gas \citep{Sanders1996}. 

The hard X-ray (up to 10 keV) imaging spectroscopy capabilities of {\em Chandra} have been particularly successful in detecting even heavily obscured pairs in the local Universe \citep{derosa_2019}. However, {\em Chandra} lacks the required sensitivity to systematically assess the AGN pair fraction among the broader population of luminous and ultra-luminous, gas-rich mergers, since sufficiently sensitive spectra cannot be obtained within moderate exposure times \citep{Iwasawa2011}. In addition, angular resolution constraints also limit the X-ray study of these systems: at $z=0.1$ {\em Chandra} can only resolve scales down to approximately 1.8 kpc.

Recent high-resolution, multi-wavelength studies have revealed several benchmark DAGN systems on sub-kiloparsec to parsec scales, offering key insights into the late stages of galaxy and BH co-evolution. In NGC~6240, a luminous infrared galaxy (LIRG) at $z = 0.024$ in an advanced merger, {\em Chandra} imaging first confirmed two obscured, hard X-ray nuclei separated by $\sim$750~pc, each showing Fe~K$\alpha$ emission \citep{komossa_2003}. Later optical spectroscopy suggested a possible third nucleus \citep{kollatschny_2020}. UGC~4211, with a $\sim$230~pc separation, was identified as a dust-obscured AGN by the wide-field infrared survey explorer (WISE) and confirmed by the Hubble space telescope (HST), Keck telescope, and Atacama large millimeter/submillimeter array (ALMA) imaging \citep{koss_2023}. Similarly, MCG--03--34--64 hosts a $\sim$100~pc dual AGN revealed by Fe~K$\alpha$ emission, where complementary HST and very large array (VLA) radio imaging provided crucial confirmation of both active nuclei \citep{falcao_2024}, underscoring the vital role of radio observations in penetrating obscured regions and distinguishing compact AGN cores in merging systems. The identification of sub-kpc separation DAGN candidates can also be effectively carried out through multi-epoch VLBI observations, such as in the case of J1543-0757 \citep[$\sim$130 pc;][]{cheng_sohn_2024}.

Local LIRGs and ultra-LIRG (ULIRGs) are especially suited candidates to identify DAGN in the nearby Universe, as they trace merging galaxies at different stages of the merging sequence. In this respect, radio observations are of paramount importance in isolating true BH accretion activity from contaminants (such as starburst, see Sec. \ref{sec:intro_DAGN}). The exceptional sensitivity and angular resolution of SKAO will allow one to carry a complete radio census of LIRGs and ULIRGs along merger sequences (early-to-late merger), allowing to bridge the gap between high-$z$ DAGN and local SMBHBs.

\subsubsection{Compact and radio-detected SMBHBs} 

Compact, gravitationally bound SMBHs that have reached or surpassed the so-called ``final parsec'' of orbital evolution \citep{begelman_1980,merritt_2013}, represent the final stage of the BH merging sequence. Despite their expected abundance \citep{volonteri_2022}, no unambiguous sub-parsec binary SMBH has yet been confirmed, largely due to the extreme angular resolution and sensitivity required to directly resolve such compact systems. Even with the state-of-the-art VLBI facilities, most searches rely on indirect signatures or are limited by the radio faintness of one component.

One of the most widely adopted strategies for identifying sub-pc SMBHBs is detecting AGN with (quasi-)periodic variations in their photometric light curves \citep{Tripathi2024}. This periodicity arises from accretion disk interactions, orbital Doppler boosting and, in the radio-band, jet precession.
Among periodic light-curve candidates, OJ~287 remains the canonical example. Its $\sim$12 yr double-peaked optical outbursts had in the past been interpreted by a variety of different models including jet precession, jet beaming due to orbital motion, or accretion-disk impact models \citep[e.g.,][]{sillanpaa_1988, valtonen_2008, Katz1997, Villata1998, Valtaoja2000}. Recent results have shown that the latest outburst of OJ 287 is dominated by non-thermal emission as predicted by jet-related outburst models \citep{Komossa2023a, Komossa2023b}, and a similar conclusion was reached for several previous outbursts \citep{Gopal2024}.
Similarly, as example, PG~1302-102 and PKS~2131-021 have shown quasi-periodic variability in the optical and radio bands, respectively \citep{Graham2015a, ONeill2022}. However, in general, distinguishing genuine periodic signals from stochastic variability (red noise) intrinsic to a single AGN remains a challenge. Recent multi-wavelength and radio spectro-photometric analyses have demonstrated that binary models for OJ~287 require a low primary mass and disk luminosity, calling for revised modeling in a new parameter regime \citep{Komossa2023a, Komossa2023b}. Currently, in order to rule out stochastic or noise-driven fluctuations, long time multi-band observations are required to confirm true periodicity \citep{Vaughan2016, ElBadry2025}. Furthermore, SMBHBs may be detected via radio-transient phenomena,  either due to temporary switch off of jet activity between the rapid SMBHBs shrinking phase and the post-coalescence new accretion disk formation \citep{Phinney2005}, or due to stellar tidal disruption event (TDE, \citealt{Liu2014}, \citealt{Shu01.2026.SKA}). Another promising approach -- enabled by radio VLBI position accuracy -- is tracing the sub-pc scale orbital motion induced by the spiraling BHs, using the radio technique of phase referencing. The 3C 66B source, one of the most studied SMBHB candidates, was initially identified with this technique, by precisely measuring the elliptical periodic offset of the (multi-band) radio-emission centroid in a Keplerian period of $\sim$1 yr, on scales of few hundreds of $\mu$as \citep{Sudou2003}. The advantage of this technique is that only one of the two BHs (preferentially the less massive) has to be radio-active.

Alternative search methods in the radio band relies on the large-scale structure morphology, such as the jets and lobes \citep{Spingola01.2026.SKA}. One is the search for helical or precessing radio jets, whose morphology may arise from orbital motion or from torque-induced jet precession in a binary system \citep{Conway1995}. Despite the difficulty in distinguishing such signatures from magnetic-field-driven helices, this technique has been applied in a few cases, such as NGC~1275 \citep{LiuChen2007}. Another technique relies on the detection of multiple episodes of jet activity, marked by multiple pairs of radio-lobes (double-double radio galaxies, DDRGs). In most DDRGs, outer lobes with steeper spectra trace past AGN activity, while inner lobes mark renewed outbursts. A rare subset, the “misaligned DDRGs”, shows jet reorientation likely caused by torques from a secondary SMBH in a non-coplanar orbit \citep{Nandi17}. VLBI monitoring reveals evolving pc-scale structures and variable radio cores \citep{Nandi24}. Signatures of a multiple BHs system can be also unveiled by the well-known S-, Z-, and X-shaped radio galaxies. In these systems, jet precession or reorientation -- driven by a secondary SMBH or disk realignment -- produces symmetric radio lobes. Such morphologies fit jet-precession models implying sub-parsec separations and are further supported by double-peaked optical emission lines \citep[e.g.,][]{Rubinur2017, Kharb2019, Misra2025}.

Finally, radio-VLBI observations currently represent the only method able to resolve the pc-scale separated SMBHBs population. As an example, the radio galaxy 0402+379 remains the most compact SMBHB candidate imaged to date. Multi-frequency very long baseline array (VLBA) observations resolved two flat-spectrum, variable radio cores separated by only 7.3 pc \citep{Rodriguez2006}, and subsequent multi-epoch VLBI astrometry over 12 years detected relative motion between them \citep{Bansal_2017}. Pc-scale and sub-pc-scale binaries have been suggested in other AGN \citep[e.g.,][]{deane2014nature, Kharb_2017}; these require further observations and monitoring.

\subsubsection{Linking to multi-messenger astronomy}

Upon final coalescence, bound binary SMBHs are the loudest sources of gravitational waves in the Universe \citep[e.g.][]{Colpi2024}; the lower-mass coalescences (chirp masses below $\sim 10^7~\mathrm{M_\odot}$) will fall in the sensitivity regime of future space-based GW interferometers like LISA, while the highest-mass coalescences (chirp masses above  10$^{8}$ M$_{\odot}$) are in the regime of PTAs. The recent detection of a GW background signal is consistent with a population of SMBHBs mergers \citep[e.g.,][]{agazie_2023, mingarelli2025}. These results are already helping constraining models for the assembly of SMBHs \citep[][]{toubiana2024, bonoli2025}. Radio observations play an invaluable role in future multi-messenger astrophysics of merging BHs. Coalescing SMBHBs would appear as radio transients (see above), once the inner disk and jet disappears shortly before coalescence, and then re-forms afterwards, and with the GW signal from coalescence in-between. Depending on the timescales and detections, any dipping radio source could represent an alert of an upcoming GW event, or an observed GW event could motivate a search for a later radio transient signal. While electromagnetic transients can also be identified in other wavebands, the radio SKAO regime is unique in combining high sensitivity and high spatial resolution, and is thus also invaluable in pinpointing the location of the counterpart (see Sec. \ref{sec:case_4}). In addition, as aforementioned, jetted TDEs that occur in evolved binary SMBH systems have characteristically different lightcurves than jetted TDEs around single SMBHs \citep{Donnarumma2015}. High-sensitivity radio observations are essential in order to follow the lightcurve evolution and model the binary orbit. As long as Swift-type X-ray transient missions operate, this approach can also be followed in X-rays (whenever the centers are not heavily absorbed), but it comes with a degeneracy that is sometimes difficult to overcome, given that both the accretion disk and jet emit strong X-ray emission, while the radio emission uniquely points to the presence of jets. This fact provides a major advantage when modeling these systems. In general, radio observations are fundamental to isolated transients signatures (such as jet precession and tidal disruption) involving the presence of SMBHBs at advanced merging stages, complementary to GW detectors. As an example, radio monitoring in combination with multi-wavelength observations of OJ 287 have also demonstrated that this system is in the LISA sensitivity regime upon coalescence \citep{Komossa2024}. Another example is the aforementioned case of 3C 66B: while the binary nature of the source has been identified with radio phase reference technique, a significant part of the parameter space of orbital solutions could already be excluded based on PTA constraints \citep{Jenet2005}. In ongoing/upcoming transient surveys at other wavelengths, such as in the optical with the Rubin large synoptic survey telescope (LSST), or in X-rays with Swift, the discovery of exceptionally large numbers of new, unidentified transients is expected \citep{Ivezic2019}. Follow-up spectroscopic observations of all of these are prohibitively time consuming. Sensitive radio observations will immediately allow us to distinguish between different source types in a way that is more challenging (or not possible) with optical and X-ray observations alone.

\subsection{Bridging the gap with simulations}

Understanding the prevalence and evolution of DAGN is central to linking galaxy mergers with BH growth. Cosmological hydrodynamic simulations and semi-analytical models of galaxy formation and evolution provide a powerful framework for exploring this connection, but they remain limited by fundamental trade-offs: large volumes are required to capture rare DAGN systems, while high resolution is needed to follow the small-scale physics of BH dynamics, accretion, and feedback. Achieving both ] is computationally expensive, and only a few recent simulations are able to resolve AGN pairs on kpc scales \citep{derosa_2019}.

DAGN, a key phase in the hierarchical assembly of galaxies, are especially sensitive to the complex interplay of subgrid processes that shape BH and galaxy dynamics both before and after the host galaxy mergers. As a result, variations in model assumptions lead to discrepancies of several orders of magnitude in the predicted number counts \citep{puerto_sanchez_2025}. These discrepancies could arise from a combination of many factors, including different selection functions, initial BH masses, BH accretion efficiency and stellar/AGN feedback prescriptions among the several simulations. Some simulations also find that the AGN feedback is so effective that it prevents the formation of DAGN in the most massive galaxies \citep{puerto_sanchez_2025, habouzit_2022}. On the one hand, the distribution of projected separations would constrain the dynamical evolution of BH pairs, revealing the physical processes that govern inward migration, the first phase of which is dynamical friction. On the other hand, the redshift distribution of DAGN will provide a direct link between the timing of BH growth and the assembly of cosmic structures.

\section{SKAO Science Case 1: Confirmation of high-$z$ DAGN candidates}
\label{sec:case_1}

The robust identification and characterization of DAGN at small ($<1$ arcsec) angular separation remains one of the major observational challenges in extragalactic astrophysics. Although promising progress has been made in the optical domain and a growing population of DAGN candidates has identified, the unambiguous confirmation of such systems requires complementary multi-wavelength observations (see Sec. \ref{sec:intro_DAGN}). In particular, since the ultimate confirmation requires spatially-resolved spectroscopy -- only available through time-expensive observations either from ground (with adaptive optics) or space observatories -- it is necessary to preliminary refine the selection only to observe the strongest candidates. This process is currently carried mainly through ground-based unresolved spectroscopy, aiming at removing the AGN-star contaminant systems that feature clear star absorption lines in the primary AGN spectra \citep{scialpi_2024}. However, depending on the separation, component flux ratio, spectra quality and galactic latitude, the efficiency of this method significantly varies from case to case. In addition, these methods can not reveal the second most important contaminant in DAGN selection, that are GLS. In this context, radio follow-ups can serve as a powerful tool to efficiently refine the selection of DAGN, helping in isolating only the strongest candidates; radio compact cores detected in correspondence of the optical centroids are an effective method to identify the DAGN nature of the candidates \citep{glikman_2023}. In addition, if multi-frequency observations are available, radio data can be used to measure the spectral index of the AGN synchrotron emission, further restricting the selection to flat-spectrum cores, although steep spectrum nuclei have also been detected especially in low luminosity AGN \citep{Giroletti2009, Panessa2013, Kharb2021}. As for the GLS contamination, we note that strong lensing is independent from the observed frequency, thus the flux ratio of the multiple images of a lensed source are expected to be the same in different bands. Then, if resolved imaging data (much time-cheaper than spectroscopy) at other bands are available, they can be coupled with radio images to exclude the lensing contaminants by comparing the flux ratio in different bands. Furthermore, multi-epoch radio observations can also be used to unveil the nature of multiple radio images: by pinpointing the radio cores and, potentially, measuring their motion over time to track their orbits (e.g., \citealt{spingola2019}), the radio band offers the most definitive method for confirming the dual nature of such systems. 

As discussed in Sec \ref{sec:intro_DAGN}, the application of VLBI to DAGN has historically been limited by flux loss and FOV constraints. By studying radio-detected ${\lesssim}$1 mJy DAGN at $z \lesssim 0.2$ with $\lesssim$3 arcsec separation, \cite{xu_2024} showed that 75\% of single AGN are undetected with VLBI, indicating that most pc-scale emission is resolved out. Although this effect is reduced at higher redshift, VLBI still probes only tens of parsecs at $0.5 < z < 3$, potentially missing larger-scale structures such as jets. 
With SKAO, the synergy between optical and radio astronomy will be fully realized. Current optical-based methods can select DAGN at $<$1 arcsec resolution down to Gaia magnitude $M_G=20.5-21$ (see Sec \ref{sec:intro_DAGN}). This means that, given the radio flux limitation imposed by VLBI with current facilities, the majority of dual systems hosting RQ AGN are ruled out from the analysis, and their characteristics are not explored yet. SKA-Mid can easily overcome these limitations. Based on the Anticipated Performance\footnote{\url{https://www.skao.int/en/science-users/122/relevant-documents}} and SKAO sensitivity calculator (SSC\footnote{\url{https://sensitivity-calculator.skao.int/}}) predictions, the most suited band for our purpose -- starting from AA* -- configuration, will be Band 5a (4.6 -- 8.5 GHz). The dense $uv$-coverage offered by SKA-Mid antennas will offer a stable sensitivity in a wide range of scales (0.35" - 21" at 6.5 GHz in configuration AA*), ensuring the full recovering of the source total flux. At the targeted redshift range ($0.5 \lesssim  z \lesssim 3$), 0.35" corresponds to ${\sim}$ 2.1 - 2.7 projected kpc, allowing to disentangle the DAGN components without resolving out the sub-kpc emission of each AGN. The resolution offered by AA* in Band 5a falls in this maximum sensitivity ``sweet spot'' range, making possible to efficiently separate the components at an angular distance comparable to that reached by current optical selection method. This will allow for the first time to investigate the RQ counterparts of optically-selected DAGN. As example, we assume a nominal target flux density of 100 $\mu$Jy at 1.4 GHz, traditionally considered the flux density at which the extragalactic radio sky starts to be dominated by RQ AGN \cite[][and reference therein]{damato_2022}. Conservatively assuming a spectral index $\alpha=-0.7$ (radio core are expected to be ``flat-spectrum'' sources), describing the synchrotron flux density at frequency $\nu$ ad $S_\nu \propto \nu^\alpha$, we derive an expected ${\sim}$35 $\mu$Jy at 6.5 GHz. Assuming a conservative 50\% fractional bandwidth, with the SSC we estimate that just in 10 min observation such an emission can be observed at ${\sim 7 \sigma}$ significance down to a resolution of 0.3", using a visibility weighting suited to balance resolution and sensitivity (Briggs robustness $= 0$). With an uniform weighting SKA-Mid AA* will reach 0.15" -- 0.05" resolution, allowing to disentangle components beyond the current optical-selection capabilities. The downgrade in sensitivity can be easily overcome: ${\sim}$2 h observations can reach the same ${\sim 7 \sigma}$ detection significance. Once AA4 will be operative, it will extend the stable sensitivity in the 0.15" - 21" range of scales, fully unlocking the SKAO potential in optically-selected DAGN follow-ups. 

The synergy between optical observatories and SKAO will not only offer an efficient way to strengthen the DAGN selection, but it will also allow to investigate their physics and interaction with the external environment. It is now well-established that the radio loudness is connected with BH accretion efficiency \citep{merloni_2003}. For single AGN, it is observed that those hosted by early-type galaxies show lower efficiency than in gas-rich systems, both in the local Universe \citep{panessa_2015} and at high redshifts \citep{damato_2022}. With a representative sample of optical-selected DAGN detected at radio frequencies we could compare the accretion efficiencies in these systems with those of single AGN. The SKAO sensitivity will allow for the first time to efficiently sample DAGN in the RQ regime (i.e. the vast majority of the population) and in the distant Universe, and put stringent constraints to the radio-detected fraction of optical emitters as a function, for example, of physical separation. Whether the brightest optical component in a DAGN also corresponds to the brightest radio emitter is not straightforward, as the two bands typically probe different accretion regimes. Hydrodynamical simulations, moreover, predict that a large fraction of the most massive BH may reside in the faintest component of the DAGN \citep{chen_2023_sim}; SKAO observations can test these predictions in the radio band down to the nJy regime, ultimately tracing differences with the optical emission of both components. 

As for the very high redshift population, the detection of DAGN way beyond the cosmic noon \citep[$z>3$,][]{mannucci_2022, yue_2023, perna_2025}, and even up to the cosmic dawn \citep[$z \gtrsim 6$,][]{yue_2021,matsuoka_2024,ubler_2024}, has started to be increasingly reported in recent years. In general, the number of radio-detected AGN at $z\gtrsim 6$ is a small fraction of the known systems, while the vast majority (${\sim}$75\%) is classified as RQ on the basis of upper limits \citep{banados_2015}. While SKAO will in general revolutionize the study of radio AGN population at the cosmic dawn, it will also enable the identification of radio counterparts of confirmed DAGN. The faintest $z \gtrsim 6$ AGN $3\sigma$ radio upper limits are of the order of $\sim 20 \mu$Jy at 1.4 GHz \citep{banados_2015}, corresponding to ${\sim}5 \mu$Jy at 6.5 GHz if $\alpha=-0.7$. As example, in just ${\sim}$50 h of AA* observations the entire population of currently known DAGN at $z>3.5$ can be observed down to a $3\sigma$ sensitivity of $\sim 0.75 \mu$Jy/beam with enough resolution to disentangle their components, immediately opening a new window on the characterization of these objects. 

Finally, we note that most of ongoing ground-based surveys searching for DAGN are carried out with adaptive optics instruments (e.g., MUSE, ERIS) mounted on 8-m class telescopes in the Southern Hemisphere \citep{mannucci_2022,mannucci_2023,scialpi_2024}. As a result, most of the identified candidates are not observable with the major radio facilities located in the Northern Hemisphere, such as VLA, VLBA, and LOFAR-VLBI. SKAO will allow for the first time to efficiently complement these ongoing and future surveys, both in confirming candidates and studying the radio properties of confirmed objects.\\

\section{SKAO Science Case 2: Radio selection of high-$z$ DAGN}
\label{sec:case_2}
The selection of DAGN in the radio band has been inefficient to date due to technical limitations of current facilities in terms of sensitivity vs. resolution combination. As an example, the deepest available VLA surveys reach few $\mu$Jy/beam noise level at 1.4 GHz, with at maximum $\sim$1.5" resolution \cite[see][and reference therein]{damato_2022}. Higher frequencies are ruled out due to the small FOV and intrinsic fainter emission. As a result, the radio-selected population of high-$z$ DAGN is currently unknown. Investigating their properties will be crucial not only to unveil the hidden obscured fraction of DAGN, but also to test many theoretical predictions. Coupled with optical follow-ups (necessary to confirm the DAGN candidate nature), the radio luminosity can be used to investigate, for example, the radio-optical displacements, radio-loudness, the relation with the BH mass and Eddington ratio ($f_{\rm Edd}$) and the accretion efficiency. A comprehensive census of the radio DAGN population will allow to discern between different evolutionary scenarios, with particular regard to the AGN feedback mode and the origin of the radio emission \citep{volonteri_2003, volonteri_2022, tremmel_2017,chen_2023_sim,shen_2023}. Current radio luminosities and fluxes of DAGN are predicted on the basis of many different prescriptions. For example, they can be driven by the BH masses in numerical simulations and \textit{assumed} accretion efficiency and merger delay, and subsequently derived from the fundamental plane \citep{merloni_2003,gultekin_2009,volonteri_2022}; otherwise, they can be inferred from the total radio power by making assumption on the accretion model efficiency and kinetic energy conversion in theoretical models \citep{meier_2001}. All these different recipes and assumptions yield several dex of uncertainty in the expected radio fluxes and in the number of DAGN that will be detected by SKAO \citep{dong_paez_2023}. 

\begin{figure}[h]
    \centering
	\includegraphics[width=\columnwidth]{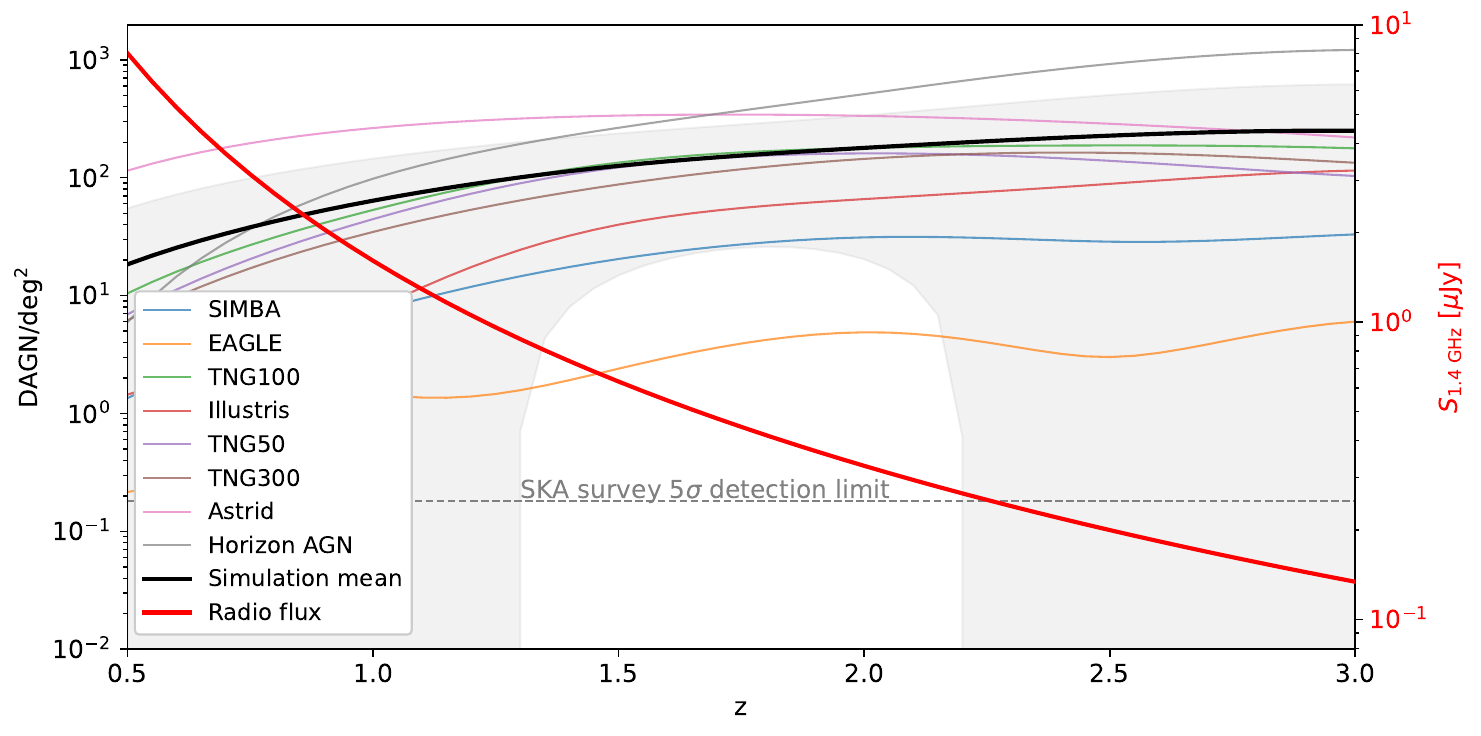}
    \caption{Number of DAGN per deg$^2$ as a function of redshift, for a number of hydrodynamical simulations. The thick black solid line is the mean of the simulations, while the gray area is the standard deviation. The thick red solid is the expected flux density at 1.4 GHz for a $L_\mathrm{bol}=10^{43}$ erg/s AGN. The grey dashed horizontal line is the $5\sigma$ detection threshold for the deep reference survey of ${\sim1}$ deg$^2$ at 1.4 GHz proposed by \cite{prandoni_2015}.}
    \label{fig:DAGN_pred}
\end{figure}

Given the aforementioned premises, we adopt as a reference the latest hydrodynamical simulation predictions presented by \cite{puerto_sanchez_2025}. In their work the number density of DAGN (nDAGN) as a function of redshift are presented for a number of hydrodynamical simulation (including Illustris, TNG50, TNG100, TNG300, EAGLE, SIMBA, Astrid, Horizon AGN; see reference therein), for different thresholds of intrinsic bolometric luminosity $L_\mathrm{bol}$ and host-galaxy stellar mass $M_\ast$. In particular, we are interested in the nDAGN of the faintest threshold adopted $L_\mathrm{bol}>10^{43}$ erg/s, for the less massive host galaxies ($M_\ast = 10^9~\mathrm{M_\odot}$). As aforementioned, to date there are no theoretical prescription to derive the radio luminosity $L_\mathrm{rad}$ from the total luminosity, and current conversions rely on empirical relations. The most studied radio correlation, both at low and high redshift, is with the X-ray emission, that is well-established for orders of magnitude of BH masses and AGN luminosities \citep{merloni_2003,panessa_2015,damato_2022}. Thus, firstly we converted the luminosity limit $L_\mathrm{bol}=10^{43}$ erg/s to X-ray luminosity, adopting for consistency the same theoretical relation reported by \citealt{puerto_sanchez_2025} (see their Eq. 3; see also \citealt{shen_2020}). Then, we converted the X-ray luminosity to $L_\mathrm{rad}$ at 1.4 GHz adopting the relation reported by \cite{damato_2022}. This relation has been derived from the so-called ``J1030'' field, that features one of the deepest joint X-ray and radio extragalactic survey to date, and is valid for RQ AGN in the $0 \lesssim z \lesssim 3$ redshift range. We note that whether these relations hold for DAGN is currently unknown, especially for the closest pairs hosted by the same galaxy. However, a study of local radio-selected DAGN showed that their X-ray properties are similar to those of single AGN \citep{gross_2019}. The luminosity has been finally converted to the observed $S_{1.4~\mathrm{GHz}}$ at 1.4 GHz, assuming $\alpha= -0.7$. 
In Fig. \ref{fig:DAGN_pred} we report the number of DAGN per deg$^2$ as a function of redshift, predicted by the several aforementioned simulations. The solid red line represents $S_{1.4~\mathrm{GHz}}$ for a $L_\mathrm{bol}=10^{43}$ erg/s source.

In \cite{prandoni_2015} a reference deep SKAO survey of ${\sim1}$ deg$^2$ at 1.4 GHz was proposed, aiming at reaching rms${\sim}50$ nJy/beam with a resolution of ${\sim}0.5$". With current AA4 specifications, the survey can be carried out in ${\sim}$1000 h. While its main scientific driver is investigating the SF and BH accretion history of the Universe, such a survey will be also of great benefit in efficiently selecting radio high-$z$ DAGN for the first time. As shown in Fig. \ref{fig:DAGN_pred}, DAGN where both the components have $L_\mathrm{bol}>10^{43}$ erg/s can be detected up to $z\sim2.2$ and down to 3-4 kpc projected separation. As a reference, current optically-based selection method can only detect DAGN with $L_\mathrm{bol}>10^{45}$ erg/s: SKAO will completely revolutionize the field,  allowing for  the first time to study both the faint and obscured DAGN populations. Importantly, it will lead to the detection of potentially hundreds of new DAGN systems at the epoch of the cosmic noon ($z\sim2$), that is where both the SF and BH accretion histories peak. For brighter sources ($L_\mathrm{bol}>10^{44}$ erg/s), the search can be extended up to $z\sim4$. We also stress that predictions reported by \cite{puerto_sanchez_2025} refer to BHs still seating in 2 distinct galaxies; thus, their predicted nDAGN can be considered lower limits.

The selection and confirmation of DAGN in the radio band has to be complemented by ancillary data. A reference survey like that proposed by \cite{prandoni_2015} will be likely followed-up by multi-wavelength photometric observations, like usually did for other deep radio surveys \citep[e.g.,][]{smolcic_2017,damato_2022}. This will, for example, enable the measurement of photometric redshifts and start formation rates of the optical counterparts, thereby allowing contaminants to be removed \citep{delvecchio_2017}. As mentioned in Sec. \ref{sec:intro_DAGN}, the $\mu$Jy radio sky at GHz frequencies is dominated by star-forming galaxies (SFG), thus complementary information from optical/infrared bands are needed to isolate the AGN emission. In addition, a maximum separation of 30 kpc would correspond to 5" -- 4" angular separation in the $0.5 < z < 3$ range; while spurious source pairs closer than this on the sky plane represent only a small fraction in current deep $\mu$Jy high-resolution surveys (e.g., 5\% -- 3\% in J1030), the increasing of source spatial density in the nJy regime can lead to a significant fraction of spurious association, if only radio separation criterion is used. Moreover, many radio sources can appear as a pair of point-like emission if only the hot-spots of the jets are detected. On the other hand, as shown by the multiwavelength characterization of the 3 GHz COSMOS survey \citep{smolcic_2017}, the radio band is crucial to unveil many low-to-moderated luminosity AGN, that are often missed by photometric classification: in a considerable fraction (16\%) of sources classified as SFG, the presence of the AGN could only be determined by the detection of the so-called ``radio-excess\footnote{Radio luminosity that is significantly higher than that expected from SF only, as inferred from standard infrared-radio relations.}''. In addition, AGN-powered radio-emission is often detected also in sources that are optically-classified as simple quiescent galaxies \citep{smolcic_2017,damato_2022}. 

Finally, we want to stress that, as mentioned in Sec. \ref{sec:intro_DAGN}, the most reliable way to confirm radio AGN emission without relying on multiwavelength follow-ups is through the measurement of the brightness temperature at low frequencies (${\sim}$150 MHz), if sub-arcsecond angular resolution is achieved \citep{morabito_2022}. Low-frequency observations have also benefits in detecting steep-spectrum emission and better isolate point-like AGN cores. In this respect, SKA-Low would in principle represent an huge step forward for studies of distant single and dual AGN. However, the lack of baselines longer than ${\sim}$80 km for SKA-Low will allow for a maximum resolution of only
${\sim}$10", insufficient to resolve sub-arcsecond DAGN. A future SKA-Low upgrade, providing at least a factor of $\sim4$ longer baselines, is necessary to fill the gap with LOFAR, which is the major current SKAO pathfinders at these frequencies, and that will implement automated acquisition and data-reduction of international baselines in the ongoing 2.0 upgrade. This upgrade would be particularly important considering that SKA-Low will be the only next-generation low-frequency instrument to have access to the entire Southern Sky, unlocking the full potential of the synergy with current and future Southern optical/infrared facilities. The lack of long baselines, however, can be mitigated by the development of a low-frequency VLBI network incorporating the SKA-Low \citep{Kobayashi01.2026.SKA}.

\section{SKAO Science Case 3: SMBHBs evolution, from wide-separation to the merger}
\label{sec:case_3}

SKAO radio observations play a fundamental role in searching for wide AGN pairs in the populations of galaxy mergers. As mentioned in Sec. \ref{sec:intro_BinBH}, a complete radio census of LIRGs and ULIRGs along the merger sequences would be a great application. These galaxies are characterized by dust-enshrouded accretion processes and, due to their merging galaxy nature, are optimal candidates to harbor merging BHs at intermediate stages between kpc-scale separated DAGN (still residing in different host galaxies) and $\lesssim1$ pc SMBHBs. The inner radio-band emission can pierce the dense inter-stellar medium, and such observations can help in characterizing the physics of the merging BHs and interaction with their host galaxies. 
Accretion onto SMBHs occurs through both radiatively efficient and inefficient processes. Each of these modes drives distinct feedback mechanisms that shape the host galaxy's evolution, and can be found in both RQ and RL AGN. 
Which is the impact --if any -- of the presence of two active BHs in shaping the properties of the final host galaxy is currently unknown (recent resolved ALMA observations of $0.4 < z < 0.8$ dual AGN suggest an increased gas fraction with respect to single AGN; \citealt{tang_2025}). SKAO will expand evolutionary studies to include the sub-$\mu$Jy RQ AGN population, while also being sensitive to the onset and earliest stages of RL AGN activity in the Universe. If intrinsic radio emission is present on pc-scales (i.e., single AGN emission is not resolved down to mas resolutions -- see caveats in Sec. \ref{fig:DAGN_census}) such as in NGC 6240 (Sec. \ref{sec:intro_BinBH}), SKA-VLBI will be able to detect \textit{and} separate single Seyfert nuclei in similar sources up to a factor of, e.g., $\times$15 in distance (for a $5\sigma$ detection, assuming the faintest NGC 6240 radio component at 1.7 GHz and 10 hour SKA-Mid Band 2 observation). SKAO exceptional sensitivity and wide survey coverage will also enable detections reaching back to the epoch of the first AGN formation (z$>7$; \citealt{prandoni_2015}). Joint with the exceptional sensitivity of JWST in the infrared band, SKAO will be able to investigate the earliest ULIRGs population, helping in isolating the SF from AGN activity. As example, REBELS-25 at $z=7.3$ is one of the highest redshift ULIRG detected by JWST \citep{hygate_2023}. With a infrared luminosity of $L_\mathrm{IR}\sim1.5\times 10^{12}~\mathrm{L_{\odot}}$ erg/s, and assuming that only 20\% of it can be ascribed to the AGN, the source would be detected at 1.4 GHz up to $z\sim8$ in future reference surveys (see Sec. \ref{sec:case_2} and \citealt{prandoni_2015}). Even if the single BH binaries could not be resolved, the SKAO-JWST synergy will extend the study of RQ/RL AGN towards the epoch of re-ionization; this will allow to investigate if, on a statistical base, merging ULIRGs have different BH-accretion properties with respect to single-galaxy AGN hosts.

Another paramount advancement that SKA-VLBI will bring, particularly important to unresolved $\lesssim 1$ pc SMBHBs science, is the exceptional astrometry accuracy that will offer \citep{paragi_2015}. As mentioned in Sec. \ref{sec:intro_BinBH}, phase reference technique has been successfully applied to measure the Keplerian period of the inspiralling BHs in  3C 66B. High astrometry precision is needed to trace the orbit on this scales, with a orbit major axis of $\sim250~\mu$as at 2.3 GHz and only $\sim50~\mu$as at 8.3 GHz (i.e., sensible to the inner jet emission). While modern VLBI interferometers can reach sufficient precision to carry out these observations, SKA-VLBI is expected to increase by an order of magnitude the astrometry precision \citep{li_2024}. This will allow to trace the smallest orbital periods in the local Universe, or sources such as 3C 66B at much higher redshift. The exceptional astrometric accuracy of SKA-VLBI will also be a fundamental tool in combination with the upcoming \textit{Gaia} Data Release 4 (DR4), which will offer multi-epoch astrometric measurement,  and thus enabling a new technique to be used to select SMBHBs.
In the radio, compact synchrotron cores trace the vicinity of accreting SMBHs, whereas optical/near-IR light is dominated by the stellar component and provides a proxy for the host's stellar-barycentric photocenter. Significant radio-optical centroid offsets are therefore expected in bound SMBH pairs with dissimilar accretion states or in post-merger recoils \citep{skipper_browne_2018}. Pinpointing such centroid offsets is challenging, requiring exquisitely precise, well-calibrated astrometry and a stable radio--optical frame tie. The SKA-Mid AA* phase will pioneer astrometric searches for binary and off-center SMBHs, offering sufficient sensitivity and resolution for targeted pilot studies. It will establish selection thresholds, quantify systematics (core shift, jet contamination, lensing), and demonstrate mas-level differential astrometry on bright, compact sources. At 5 GHz, AA* provides $\sim$0.3" beams, corresponding to $\sim$0.6–2.6 kpc over $z=0.1-4$. In AA*+VLBI mode, phased SKA-Mid can form four simultaneous VLBI beams within its primary field, enabling true in-field multi-view calibration. This suppresses atmospheric and geometric errors, achieving sub-mas relative astrometry and allowing sub-kpc displacement tests, with tens-of-parsec resolution over $z\approx0.5-2$. The SKA-Mid AA4 phase will deliver $\sim$80 mas beams at 5 GHz, directly resolving $\sim$0.15–0.6 kpc over $z=0.1-4$. In AA4+VLBI, Earth-scale baselines yield mas–sub-mas resolution and $\lesssim$mas differential astrometry, reaching tens of $\mu$as for bright, compact cores (1 mas $\sim$ 6–7 pc at $z=0.5-4$). Combined with Euclid and Rubin/LSST host and variability data, and \textit{Gaia} DR4 epoch astrometry, these will enable the first statistically robust sample and demographic constraints on binary and recoiling SMBHs in the pc–kpc regime. However, we note that at centimeter wavelengths the central engine may be physically offset from the radio core, by up to $10^3-10^5$ gravitational radii along the jet \citep{lobanov_1998,marscher_2008}. This potential displacement must be accounted for when interpreting radio-optical centroid offsets in candidate binary systems.

The analysis of transient phenomena will be fundamental to investigate the physics of the final stages of SMBHBs merges, i.e. the sources of GW that will be detected by LISA and PTAs (see Sec. \ref{sec:intro_BinBH}). The SKAO will detect the progenitors of the merging BHs (with $\sim 10^{4}-10^{7}\, \rm M_{\odot}$) that LISA will later characterize. Together, these two observatories will provide a comprehensive view of the full evolutionary pathway from galaxy mergers to BH coalescence, as well as the physical mechanisms driving BH accretion throughout galaxy mergers. Furthermore, the population of the most massive SMBHBs ($M_{\rm BH}\gtrsim 10^9 ~\mathrm{M_\odot}$) will be directly constrained with increasing precision using the SKA-Mid and PTAs \citep{Shannon01.2026.SKA,Takahashi01.2026.SKA}. \\
Both jetted TDE and disk accretion disruption (and post-coalescence re-formation) events are expected to happen temporarily close to the merging event, and they would produce a significant variation in the observed radio flux. In the LISA frequency and sensibility range, GW source can be started to be detected days or months before the coalescence, depending on the BHs mass and redshift. As example, a SMBHBs with total mass of $10^{6}~\mathrm{M_\odot}$ at $z=3$ will be detected by LISA approximately 1 month before merging \citep{sesana_2021}. However, also depending on the SMBHBs mass and redshift, the astrometric uncertainty of the event can be up to several hundreds and square deg, reducing to a few square deg $\sim$1 day before merging \citep[e.g.,][]{mangiagli_2020}. In this frame, SKAO will be of great value in identify the upcoming GW counterpart \citep{GemmaAnderson01.2026.SKA}. As example, a proposed wide-area reference survey will cover all the observable sky ($\sim$31000 $\mathrm{deg^2}$) at 1.4 GHz, down to 5 $\mu$Jy/beam sensitivity. When a GW event starts to be detected by LISA, a monitoring of the corresponding area can be triggered with SKAO. Thanks to the unique combination of sensitivity and large FOV (1 $\mathrm{deg^2}$ at 1.4 GHz), SKAO can cover a large area around the LISA detected event down to comparable sensitivity of the reference survey in few tens hour observations, allowing for multiple scans before the final coalescence. Any switched-off source with respect to the reference survey flux will emerge as a strong candidate for the GW source. On the other hand, the unique astrometric precision, resolution and sensitivity of SKA-VLBI will allow to identify the precursor population of massive GW emitters (see above and Sec. \ref{sec:intro_BinBH}), when the BHs are still spiraling at (sub-)pc scale separations, ultimately covering all the stages of SMBHs mergers. 

Finally, as mentioned in Sec. \ref{sec:case_1}, SKAO will offer an unmatched image quality and stable sensitivity in a wide range of angular scales at all observing frequencies. Observing simulations have shown how both SKA-Mid and SKA-Low will dramatically increase the image quality with respect to, e.g., VLBA and LOFAR, even in simple snapshot observations and already with AA* configuration \citep{braun2019anticipatedperformancesquarekilometre}. The possibility to precisely model the multi-frequency emission of extended and complex sources, from the mas to the degree scales, will revolutionize the field of S-, X- and Z-shaped radio-galaxies and DDRGs (Sec. \ref{sec:intro_BinBH}). In this respect, the combination of SKA-Mid, SKA-Low and VLBI represent an unique synergy that will allow to distinguish signatures of sub-pc SMBHs pairs, such as the presence of jet precession, by fitting the spectral index maps of these sources with an unprecedented sensitivity and image quality at all physical scales \citep{Rubinur2017,Kharb2019}.

\section{SKAO Science Case 4: Testing simulation predictions}
\label{sec:case_4}

\begin{figure}[h]
    \centering
	\includegraphics[width=\columnwidth]{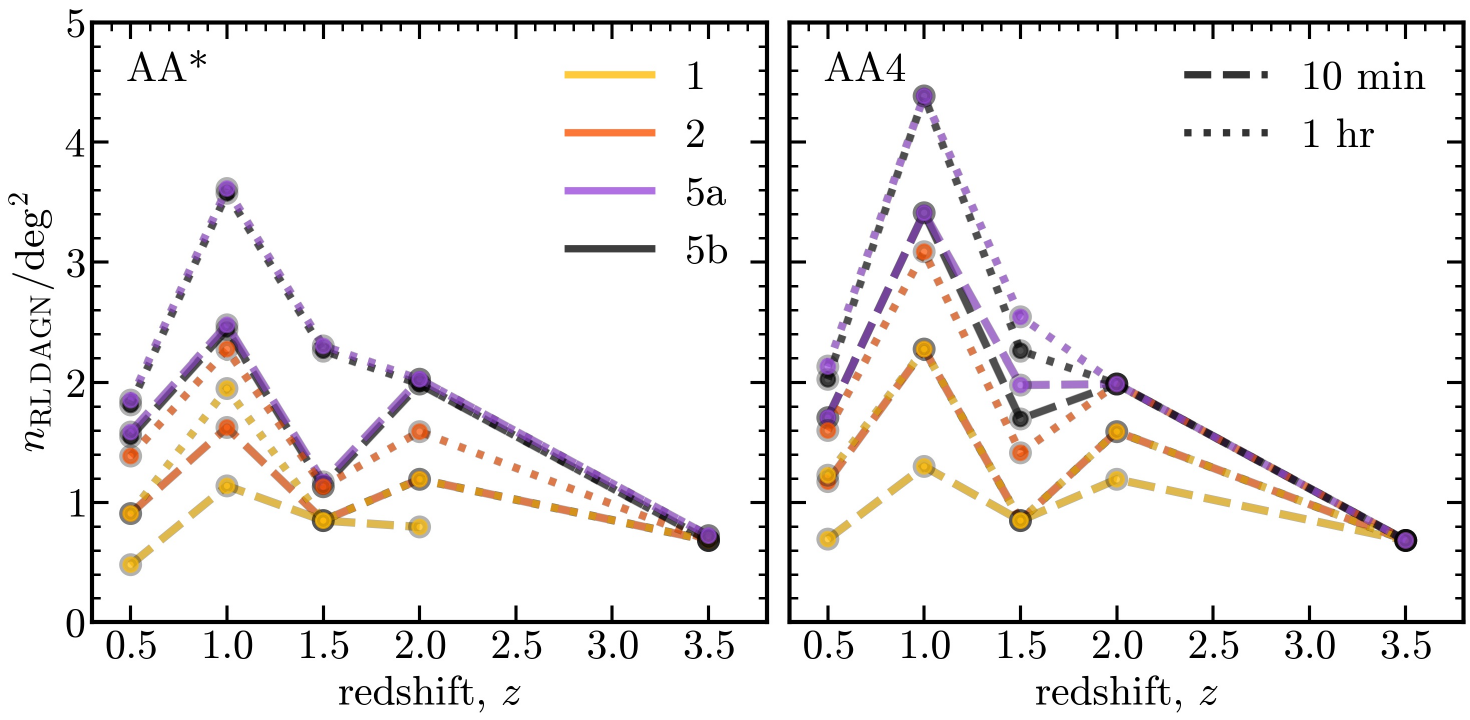}
    \caption{Predicted sky density of RL DAGN per square degree identifiable with SKA AA* (left) and AA4 (right), based on the SIMBA cosmological hydrodynamical simulations. Lines correspond to observation times of 10 minutes (dashed) and 1 hour (dotted) for SKA Bands 1 (yellow), 2 (orange), 5a (purple), and 5b (black). Figure adapted from \citet{pillay_inprep}.}
    \label{fig:simba-predict}
\end{figure}

Ahead of the SKAO era, cosmological simulations, which previously could not be compared in detail to observations due to technical limitations, will play a key role in identifying and characterizing AGN populations.

For instance, \citet{thomas_2024} compared the radio excess galaxy population in SIMBA simulation with the MeerKAT international GHz tiered extragalactic exploration (MIGHTEE) survey \citep{jarvis_2017, heywood_2022, whittam_2022}, examining their properties as functions of redshift and radio luminosity. They found broad agreement in the overall demographics, such as the similarity in the $M_\ast$ and $f_{\rm Edd}$ functions for high-excitation and low excitation radio galaxies (HERGs and LERGs, respectively), and confirm the presence of the predicted inefficiently accreting HERG populations. Discrepancies were also uncovered as the HERG–LERG distinction in SIMBA simulation was found to be more pronounced due to differences in host galaxy gas content. With its superior sensitivity and angular resolution, SKAO will extend such comparisons to the DAGN regime, offering a diverse set of powerful tests of physical models in cosmological simulations.

Furthermore, these simulations provide a framework to forecast the DAGN population potentially observable with the SKAO, as seen in Sec. ~\ref{sec:case_2}. While Fig. \ref{fig:DAGN_pred} provides a valuable reference to compare current large-scale simulations, \citet{pillay_inprep} take a complementary approach, evaluating whether each simulated DAGN would be detectable and spatially resolvable with the projected SKAO capabilities; that is, whether both AGN are bright enough to be detected and sufficiently separated to be resolved.

Following \citet{thomas_2021, pillay_inprep}, we adopt the \citet{kording_2008} prescription for the radio emission of SIMBA's DAGN population within 30 pkpc with $M_{\rm BH} \geq 10^8 \rm \, M_{\odot}$ and $0 < f_{\rm Edd} < 0.02$ hosted in galaxies with $M_\ast \geq 10^{9.5} \rm \, M_{\odot}$. The result radio flux density estimates are compared to SKA-Mid sensitivity limits from the current SSC for a hypothetical observation at RA$=0\degree$, Declination $=-45\degree$, and a maximum elevation of $59.2\degree$ (using resolution-sensitivity balanced visibility weighting corresponding to Briggs robustness $= 0$).

Figure~\ref{fig:simba-predict} shows the counts per square degree of RL DAGN predicted to be identifiable with AA4 for a 10-minute and 1-hour observation time; SIMBA predicts up to $\sim 10$ RL DAGN per square degree. An SKAO survey of $\sim7000$ deg$^2$ could thus detect on the order $10^{3-4}$ RL DAGN, based on these predictions. These results indicate that the all of the simulated systems that are bright enough to be detected are also spatially resolvable. Thus, the effective limit on what can be observed is set by the faintness of the DAGN population and the sensitivity of SKAO (or depth of a given survey). These results provide a basis for optimizing future SKAO survey strategies by evaluating outcomes across different frequency bands and integration times. This is particularly true when one considers synoptic survey strategies that would reveal both single and dual AGN candidates through flux density variability. 
 
Based on these predictions, the SKA-Mid is well-poised to deliver a statistically significant census of RL DAGN, enabling direct constraints on their abundance, demographics, and evolution.

\section{Conclusions and final remarks}
In this chapter, we investigated the critical role of the SKAO in achieving a comprehensive understanding of DAGN and SMBHBs. We showed that the study of these multi-SMBH systems is paramount for constraining models of hierarchical galaxy assembly and the co-evolution of SMBHs with their hosts. We highlighted that while radio emission provides a unique, dust-unbiased tool for identifying and confirming multiple accreting SMBHs, current limitations (like insufficient sensitivity and resolution combination, and small area coverage) have hindered the detection of statistically robust samples, particularly for faint, high-$z$, or obscured sources. SKAO will overcome these constraints through its unmatched combination of sensitivity, angular resolution, image fidelity and survey speed.

We analyzed how SKA-Mid will refine the selection of high-$z$ DAGN candidates previously identified through selection techniques at other wavelengths (mainly optical). Specifically, we detailed how the stable sensitivity and resolution provided by the AA* configuration in Band 5a will enable us to efficiently confirm the DAGN nature of these candidates, exclude contaminants like GLS, and crucially, explore the characteristics of the faint RQ population currently missed by traditional VLBI surveys. Furthermore, we demonstrated that deep reference surveys proposed for SKAO will allow for the blind selection of obscured DAGN. We showed that SKAO will reach the sensitivity necessary (down to nJy/beam levels) to detect systems where both components have bolometric luminosities $L_\mathrm{bol}>10^{43}$ erg/s up to $z\sim2.2$. his capability represents a revolutionary extension beyond current optical limits ($L_\mathrm{bol}>10^{45}$ erg/s), promising the detection of potentially hundreds of new DAGN systems during the epoch of cosmic noon.

We investigated the potential for SKAO to trace the entire SMBHBs merger sequence. This includes studying wide-separation pairs in gas-rich mergers like LIRGs and ULIRGs from the local Universe to the cosmic dawn, where radio observation is essential for distinguishing accretion activity from SF. For compact, sub-parsec SMBHBs, we explored indirect signatures, such as jet precession, which generate characteristic large-scale structures like DDRGs and S-, X-, and Z-shaped radio galaxies. Most importantly, we emphasized the critical role of SKA-VLBI in achieving an order-of-magnitude increase in astrometry precision. This will enable us to trace sub-parsec orbital motions via phase referencing (like, e.g., for the binary 3C 66B) and utilize the highly accurate radio-optical centroid offsets—pioneered with SKA-Mid AA* and refined with AA4, to deliver the first statistically robust constraints on binary and recoiling SMBHs in the pc–kpc regime. However, we point out that, despite the unprecedented capabilities of the SKAO, significant interpretative challenges remain, such as those related to the AGN duty cycle. This can prevent the detection of dual radio cores, where the identification of only a single core in a DAGN can result from a temporarily inactive AGN. At high redshift, in particular, AGN variability can cause the secondary component to be too faint, falling below the sensitivity limit. Nevertheless, we note that in the radio band it may still be possible to detect relic or extended emission (e.g., from older jet activity) even if the core itself is not currently radio-active. High redshift DAGN research can also be affected by foreground contaminants, where the presence of compact radio cores at close projected separation can significantly impact the efficiency of DAGN selection. Disentangling true AGN pairs from projection effects, modeling complex jet structures in low-luminosity or high-redshift environments, and accounting for AGN duty cycles and foreground contamination will require sophisticated modeling and deep multi-wavelength follow-ups to minimize biases.

We established that SKAO is integral to future multi-messenger astrophysics by identifying the electromagnetic counterparts to low-frequency GW events detected by LISA. Its ability to conduct rapid, wide-and-fast monitoring is necessary to pinpoint radio transients, such as temporary jet switch-offs. Finally, we concluded that the demographics and evolution of RL DAGN derived from SKAO observations will provide crucial constraints for cosmological hydrodynamic simulations, moving the field toward observationally driven models of SMBH dynamics and feedback across cosmic time.

\bibliographystyle{abbrvnat-maxbibnames4}
\bibliography{chapter} 

@ARTICLE{mangiagli_2020,
       author = {{Mangiagli}, Alberto and {Klein}, Antoine and {Bonetti}, Matteo and {Katz}, Michael L. and {Sesana}, Alberto and {Volonteri}, Marta and {Colpi}, Monica and {Marsat}, Sylvain and {Babak}, Stanislav},
        title = "{Observing the inspiral of coalescing massive black hole binaries with LISA in the era of multimessenger astrophysics}",
      journal = {\prd},
         year = 2020,
        month = oct,
       volume = {102},
       number = {8},
          eid = {084056},
        pages = {084056},
          doi = {10.1103/PhysRevD.102.084056},
archivePrefix = {arXiv},
       eprint = {2006.12513},
 primaryClass = {astro-ph.HE},
       adsurl = {https://ui.adsabs.harvard.edu/abs/2020PhRvD.102h4056M}
}

@ARTICLE{cheng_sohn_2024,
       author = {{Cheng}, Xiaopeng and {Sohn}, Bong Won},
        title = "{Two Radio Cores in GPS J1543-0757: A New Dual Supermassive Black Hole System?}",
      journal = {\apj},
         year = 2024,
        month = oct,
       volume = {974},
       number = {2},
          eid = {155},
        pages = {155},
          doi = {10.3847/1538-4357/ad6df9},
archivePrefix = {arXiv},
       eprint = {2408.06187},
 primaryClass = {astro-ph.GA},
       adsurl = {https://ui.adsabs.harvard.edu/abs/2024ApJ...974..155C}
}

@ARTICLE{Katz1997,
       author = {{Katz}, J.~I.},
        title = "{A Precessing Disk in OJ 287?}",
      journal = {\apj},
         year = 1997,
        month = mar,
       volume = {478},
       number = {2},
        pages = {527-529},
          doi = {10.1086/303811},
       adsurl = {https://ui.adsabs.harvard.edu/abs/1997ApJ...478..527K}
}

@ARTICLE{Villata1998,
       author = {{Villata}, M. and {Raiteri}, C.~M. and {Sillanpaa}, A. and {Takalo}, L.~O.},
        title = "{A beaming model for the OJ 287 periodic optical outbursts}",
      journal = {\mnras},
         year = 1998,
        month = jan,
       volume = {293},
       number = {1},
        pages = {L13-L16},
          doi = {10.1046/j.1365-8711.1998.01244.x},
       adsurl = {https://ui.adsabs.harvard.edu/abs/1998MNRAS.293L..13V}
}

@ARTICLE{Gopal2024,
       author = {{Gopal-Krishna}},
        title = "{Clues on the nature of the quasi-periodic optical outbursts of the blazar OJ 287}",
      journal = {\aap},
         year = 2024,
        month = aug,
       volume = {688},
          eid = {L16},
        pages = {L16},
          doi = {10.1051/0004-6361/202449409},
archivePrefix = {arXiv},
       eprint = {2407.09273},
 primaryClass = {astro-ph.HE},
       adsurl = {https://ui.adsabs.harvard.edu/abs/2024A&A...688L..16G}
}

@ARTICLE{terashima_2003,
       author = {{Terashima}, Yuichi and {Wilson}, Andrew S.},
        title = "{Chandra Snapshot Observations of Low-Luminosity Active Galactic Nuclei with a Compact Radio Source}",
      journal = {\apj},
         year = 2003,
        month = jan,
       volume = {583},
       number = {1},
        pages = {145-158},
          doi = {10.1086/345339},
archivePrefix = {arXiv},
       eprint = {astro-ph/0209607},
 primaryClass = {astro-ph},
       adsurl = {https://ui.adsabs.harvard.edu/abs/2003ApJ...583..145T}
}

@ARTICLE{Valtaoja2000,
       author = {{Valtaoja}, E. and {Ter{\"a}sranta}, H. and {Tornikoski}, M. and {Sillanp{\"a}{\"a}}, A. and {Aller}, M.~F. and {Aller}, H.~D. and {Hughes}, P.~A.},
        title = "{Radio Monitoring of OJ 287 and Binary Black Hole Models for Periodic Outbursts}",
      journal = {\apj},
         year = 2000,
        month = mar,
       volume = {531},
       number = {2},
        pages = {744-755},
          doi = {10.1086/308494},
       adsurl = {https://ui.adsabs.harvard.edu/abs/2000ApJ...531..744V}
}

@ARTICLE{Komossa2024,
       author = {{Komossa}, S. and {Grupe}, D.},
        title = "{The Extremes of Continuum and Emission-Line Variability of AGN: Changing-Look Events and Binary SMBHS}",
      journal = {Serbian Astronomical Journal},
         year = 2024,
        month = dec,
       volume = {209},
        pages = {1-24},
          doi = {10.2298/SAJ2409001K},
       adsurl = {https://ui.adsabs.harvard.edu/abs/2024SerAJ.209....1K}
}

@ARTICLE{lobanov_1998,
       author = {{Lobanov}, A.~P.},
        title = "{Ultracompact jets in active galactic nuclei}",
      journal = {\aap},
         year = 1998,
        month = feb,
       volume = {330},
        pages = {79-89},
          doi = {10.48550/arXiv.astro-ph/9712132},
archivePrefix = {arXiv},
       eprint = {astro-ph/9712132},
 primaryClass = {astro-ph},
       adsurl = {https://ui.adsabs.harvard.edu/abs/1998A&A...330...79L}
}

@ARTICLE{marscher_2008,
       author = {{Marscher}, Alan P. and {Jorstad}, Svetlana G. and {D'Arcangelo}, Francesca D. and {Smith}, Paul S. and {Williams}, G. Grant and {Larionov}, Valeri M. and {Oh}, Haruki and {Olmstead}, Alice R. and {Aller}, Margo F. and {Aller}, Hugh D. and {McHardy}, Ian M. and {L{\"a}hteenm{\"a}ki}, Anne and {Tornikoski}, Merja and {Valtaoja}, Esko and {Hagen-Thorn}, Vladimir A. and {Kopatskaya}, Eugenia N. and {Gear}, Walter K. and {Tosti}, Gino and {Kurtanidze}, Omar and {Nikolashvili}, Maria and {Sigua}, Lorand and {Miller}, H. Richard and {Ryle}, Wesley T.},
        title = "{The inner jet of an active galactic nucleus as revealed by a radio-to-{\ensuremath{\gamma}}-ray outburst}",
      journal = {\nat},
         year = 2008,
        month = apr,
       volume = {452},
       number = {7190},
        pages = {966-969},
          doi = {10.1038/nature06895},
       adsurl = {https://ui.adsabs.harvard.edu/abs/2008Natur.452..966M}
}

@ARTICLE{bonzini_2013,
       author = {{Bonzini}, M. and {Padovani}, P. and {Mainieri}, V. and {Kellermann}, K.~I. and {Miller}, N. and {Rosati}, P. and {Tozzi}, P. and {Vattakunnel}, S.},
        title = "{The sub-mJy radio sky in the Extended Chandra Deep Field-South: source population}",
      journal = {\mnras},
         year = 2013,
        month = dec,
       volume = {436},
       number = {4},
        pages = {3759-3771},
          doi = {10.1093/mnras/stt1879},
archivePrefix = {arXiv},
       eprint = {1310.1248},
 primaryClass = {astro-ph.CO},
       adsurl = {https://ui.adsabs.harvard.edu/abs/2013MNRAS.436.3759B}
}

@ARTICLE{delvecchio_2017,
       author = {{Delvecchio}, I. and {Smol{\v{c}}i{\'c}}, V. and {Zamorani}, G. and {Lagos}, C. Del P. and {Berta}, S. and {Delhaize}, J. and {Baran}, N. and {Alexander}, D.~M. and {Rosario}, D.~J. and {Gonzalez-Perez}, V. and {Ilbert}, O. and {Lacey}, C.~G. and {Le F{\`e}vre}, O. and {Miettinen}, O. and {Aravena}, M. and {Bondi}, M. and {Carilli}, C. and {Ciliegi}, P. and {Mooley}, K. and {Novak}, M. and {Schinnerer}, E. and {Capak}, P. and {Civano}, F. and {Fanidakis}, N. and {Herrera Ruiz}, N. and {Karim}, A. and {Laigle}, C. and {Marchesi}, S. and {McCracken}, H.~J. and {Middleberg}, E. and {Salvato}, M. and {Tasca}, L.},
        title = "{The VLA-COSMOS 3 GHz Large Project: AGN and host-galaxy properties out to z {\ensuremath{\lesssim}} 6}",
      journal = {\aap},
         year = 2017,
        month = jun,
       volume = {602},
          eid = {A3},
        pages = {A3},
          doi = {10.1051/0004-6361/201629367},
archivePrefix = {arXiv},
       eprint = {1703.09720},
 primaryClass = {astro-ph.GA},
       adsurl = {https://ui.adsabs.harvard.edu/abs/2017A&A...602A...3D}
}

@ARTICLE{DiMatteo2005,
       author = {{Di Matteo}, Tiziana and {Springel}, Volker and {Hernquist}, Lars},
        title = "{Energy input from quasars regulates the growth and activity of black holes and their host galaxies}",
      journal = {\nat},
         year = 2005,
        month = feb,
       volume = {433},
       number = {7026},
        pages = {604-607},
          doi = {10.1038/nature03335},
archivePrefix = {arXiv},
       eprint = {astro-ph/0502199},
 primaryClass = {astro-ph},
       adsurl = {https://ui.adsabs.harvard.edu/abs/2005Natur.433..604D}
}

@ARTICLE{Colpi2024,
       author = {{Colpi}, Monica and {Danzmann}, Karsten and {Hewitson}, Martin and {Holley-Bockelmann}, Kelly and {Jetzer}, Philippe and {Nelemans}, Gijs and {Petiteau}, Antoine and {Shoemaker}, David and {Sopuerta}, Carlos and {Stebbins}, Robin and {Tanvir}, Nial and {Ward}, Henry and {Weber}, William Joseph and {Thorpe}, Ira and {Daurskikh}, Anna and {Deep}, Atul and {Fern{\'a}ndez N{\'u}{\~n}ez}, Ignacio and {Garc{\'\i}a Marirrodriga}, C{\'e}sar and {Gehler}, Martin and {Halain}, Jean-Philippe and {Jennrich}, Oliver and {Lammers}, Uwe and {Larra{\~n}aga}, Jonan and {Lieser}, Maike and {L{\"u}tzgendorf}, Nora and {Martens}, Waldemar and {Mondin}, Linda and {Piris Ni{\~n}o}, Ana and {Amaro-Seoane}, Pau and {Arca Sedda}, Manuel and {Auclair}, Pierre and {Babak}, Stanislav and {Baghi}, Quentin and {Baibhav}, Vishal and {Baker}, Tessa and {Bayle}, Jean-Baptiste and {Berry}, Christopher and {Berti}, Emanuele and {Boileau}, Guillaume and {Bonetti}, Matteo and {Brito}, Richard and {Buscicchio}, Riccardo and {Calcagni}, Gianluca and {Capelo}, Pedro R. and {Caprini}, Chiara and {Caputo}, Andrea and {Castelli}, Eleonora and {Chen}, Hsin-Yu and {Chen}, Xian and {Chua}, Alvin and {Davies}, Gareth and {Derdzinski}, Andrea and {Domcke}, Valerie Fiona and {Doneva}, Daniela and {Dvorkin}, Irna and {Mar{\'\i}a Ezquiaga}, Jose and {Gair}, Jonathan and {Haiman}, Zoltan and {Harry}, Ian and {Hartwig}, Olaf and {Hees}, Aurelien and {Heffernan}, Anna and {Husa}, Sascha and {Izquierdo-Villalba}, David and {Karnesis}, Nikolaos and {Klein}, Antoine and {Korol}, Valeriya and {Korsakova}, Natalia and {Kupfer}, Thomas and {Laghi}, Danny and {Lamberts}, Astrid and {Larson}, Shane and {Le Jeune}, Maude and {Lewicki}, Marek and {Littenberg}, Tyson and {Madge}, Eric and {Mangiagli}, Alberto and {Marsat}, Sylvain and {Vilchez}, Ivan Martin and {Maselli}, Andrea and {Mathews}, Josh and {van de Meent}, Maarten and {Muratore}, Martina and {Nardini}, Germano and {Pani}, Paolo and {Peloso}, Marco and {Pieroni}, Mauro and {Pound}, Adam and {Quelquejay-Leclere}, Hippolyte and {Ricciardone}, Angelo and {Rossi}, Elena Maria and {Sartirana}, Andrea and {Savalle}, Etienne and {Sberna}, Laura and {Sesana}, Alberto and {Shoemaker}, Deirdre and {Slutsky}, Jacob and {Sotiriou}, Thomas and {Speri}, Lorenzo and {Staab}, Martin and {Steer}, Dani{\`e}le and {Tamanini}, Nicola and {Tasinato}, Gianmassimo and {Torrado}, Jesus and {Torres-Orjuela}, Alejandro and {Toubiana}, Alexandre and {Vallisneri}, Michele and {Vecchio}, Alberto and {Volonteri}, Marta and {Yagi}, Kent and {Zwick}, Lorenz},
        title = "{LISA Definition Study Report}",
      journal = {arXiv e-prints},
         year = 2024,
        month = feb,
          eid = {arXiv:2402.07571},
        pages = {arXiv:2402.07571},
          doi = {10.48550/arXiv.2402.07571},
archivePrefix = {arXiv},
       eprint = {2402.07571},
 primaryClass = {astro-ph.CO},
       adsurl = {https://ui.adsabs.harvard.edu/abs/2024arXiv240207571C}
}

@ARTICLE{Tripathi2024,
       author = {{Tripathi}, Ashutosh and {Gupta}, Alok C. and {Smith}, Krista Lynne and {Wiita}, Paul J. and {Aller}, Margo F. and {Volvach}, Alexandr E. and {L{\"a}hteenm{\"a}ki}, Anne and {Aller}, Hugh D. and {Tornikoski}, Merja and {Volvach}, Larisa N.},
        title = "{Revisiting Radio Variability of the Blazar 3C 454.3}",
      journal = {\apj},
         year = 2024,
        month = dec,
       volume = {977},
       number = {2},
          eid = {166},
        pages = {166},
          doi = {10.3847/1538-4357/ad90e3},
archivePrefix = {arXiv},
       eprint = {2412.10771},
 primaryClass = {astro-ph.HE},
       adsurl = {https://ui.adsabs.harvard.edu/abs/2024ApJ...977..166T}
}

@ARTICLE{ONeill2022,
       author = {{O'Neill}, S. and {Kiehlmann}, S. and {Readhead}, A.~C.~S. and {Aller}, M.~F. and {Blandford}, R.~D. and {Liodakis}, I. and {Lister}, M.~L. and {Mr{\'o}z}, P. and {O'Dea}, C.~P. and {Pearson}, T.~J. and {Ravi}, V. and {Vallisneri}, M. and {Cleary}, K.~A. and {Graham}, M.~J. and {Grainge}, K.~J.~B. and {Hodges}, M.~W. and {Hovatta}, T. and {L{\"a}hteenm{\"a}ki}, A. and {Lamb}, J.~W. and {Lazio}, T.~J.~W. and {Max-Moerbeck}, W. and {Pavlidou}, V. and {Prince}, T.~A. and {Reeves}, R.~A. and {Tornikoski}, M. and {Vergara de la Parra}, P. and {Zensus}, J.~A.},
        title = "{The Unanticipated Phenomenology of the Blazar PKS 2131-021: A Unique Supermassive Black Hole Binary Candidate}",
      journal = {\apjl},
         year = 2022,
        month = feb,
       volume = {926},
       number = {2},
          eid = {L35},
        pages = {L35},
          doi = {10.3847/2041-8213/ac504b},
archivePrefix = {arXiv},
       eprint = {2111.02436},
 primaryClass = {astro-ph.HE},
       adsurl = {https://ui.adsabs.harvard.edu/abs/2022ApJ...926L..35O}
}

@ARTICLE{Komossa2023a,
       author = {{Komossa}, S. and {Grupe}, D. and {Kraus}, A. and {Gurwell}, M.~A. and {Haiman}, Z. and {Liu}, F.~K. and {Tchekhovskoy}, A. and {Gallo}, L.~C. and {Berton}, M. and {Blandford}, R. and {G{\'o}mez}, J.~L. and {Gonzalez}, A.~G.},
        title = "{Absence of the predicted 2022 October outburst of OJ 287 and implications for binary SMBH scenarios}",
      journal = {\mnras},
         year = 2023,
        month = jun,
       volume = {522},
       number = {1},
        pages = {L84-L88},
          doi = {10.1093/mnrasl/slad016},
archivePrefix = {arXiv},
       eprint = {2302.11646},
 primaryClass = {astro-ph.HE},
       adsurl = {https://ui.adsabs.harvard.edu/abs/2023MNRAS.522L..84K}
}

@ARTICLE{Komossa2023b,
       author = {{Komossa}, S. and {Kraus}, A. and {Grupe}, D. and {Gonzalez}, A.~G. and {Gurwell}, M.~A. and {Gallo}, L.~C. and {Liu}, F.~K. and {Myserlis}, I. and {Krichbaum}, T.~P. and {Laine}, S. and {Bach}, U. and {G{\'o}mez}, J.~L. and {Parker}, M.~L. and {Yao}, S. and {Berton}, M.},
        title = "{MOMO. VI. Multifrequency Radio Variability of the Blazar OJ 287 from 2015 to 2022, Absence of Predicted 2021 Precursor-flare Activity, and a New Binary Interpretation of the 2016/2017 Outburst}",
      journal = {\apj},
         year = 2023,
        month = feb,
       volume = {944},
       number = {2},
          eid = {177},
        pages = {177},
          doi = {10.3847/1538-4357/acaf71},
archivePrefix = {arXiv},
       eprint = {2302.11486},
 primaryClass = {astro-ph.HE},
       adsurl = {https://ui.adsabs.harvard.edu/abs/2023ApJ...944..177K}
}

@ARTICLE{Mingarelli2025,
       author = {{Mingarelli}, C.~M.~F. and {Blecha}, L. and {Bogdanovi{\'c}}, T. and {Charisi}, M. and {Chen}, S. and {Escala}, A. and {Goncharov}, B. and {Graham}, M.~J. and {Komossa}, S. and {McWilliams}, S.~T. and {Schwartz}, D.~A. and {Zrake}, J.},
        title = "{Insights into supermassive black hole mergers from the gravitational wave background}",
      journal = {Nature Astronomy},
         year = 2025,
        month = feb,
       volume = {9},
        pages = {183-184},
          doi = {10.1038/s41550-025-02482-1},
archivePrefix = {arXiv},
       eprint = {2501.08956},
 primaryClass = {astro-ph.HE},
       adsurl = {https://ui.adsabs.harvard.edu/abs/2025NatAs...9..183M}
}

@ARTICLE{Misra2025,
       author = {{Misra}, Arpita and {Jamrozy}, Marek and {We{\.z}gowiec}, Marek and {Kozie{\l}-Wierzbowska}, Dorota},
        title = "{Multiwavelength investigations of PKS 2300{\textendash}18: S-shaped radio quasar with precessing jets and double-peaked broad emission-line spectrum}",
      journal = {\mnras},
         year = 2025,
        month = jan,
       volume = {536},
       number = {3},
        pages = {2025-2045},
          doi = {10.1093/mnras/stae2639},
archivePrefix = {arXiv},
       eprint = {2502.09441},
 primaryClass = {astro-ph.GA},
       adsurl = {https://ui.adsabs.harvard.edu/abs/2025MNRAS.536.2025M}
}

@ARTICLE{Rubinur2017,
       author = {{Rubinur}, K. and {Das}, M. and {Kharb}, P. and {Honey}, M.},
        title = "{A candidate dual AGN in a double-peaked emission-line galaxy with precessing radio jets}",
      journal = {\mnras},
         year = 2017,
        month = mar,
       volume = {465},
       number = {4},
        pages = {4772-4782},
          doi = {10.1093/mnras/stw2981},
       adsurl = {https://ui.adsabs.harvard.edu/abs/2017MNRAS.465.4772R}
}

@ARTICLE{Rubinur2019,
       author = {{Rubinur}, K. and {Das}, M. and {Kharb}, P.},
        title = "{Searching for dual AGN in galaxies with double-peaked emission line spectra using radio observations}",
      journal = {\mnras},
         year = 2019,
        month = apr,
       volume = {484},
       number = {4},
        pages = {4933-4950},
          doi = {10.1093/mnras/stz334},
archivePrefix = {arXiv},
       eprint = {1902.00689},
 primaryClass = {astro-ph.GA},
       adsurl = {https://ui.adsabs.harvard.edu/abs/2019MNRAS.484.4933R}
}

@ARTICLE{Kharb2019,
       author = {{Kharb}, P. and {Vaddi}, S. and {Sebastian}, B. and {Subramanian}, S. and {Das}, M. and {Paragi}, Z.},
        title = "{A Curved 150 pc Long Jet in the Double-peaked Emission-line AGN KISSR 434}",
      journal = {\apj},
         year = 2019,
        month = feb,
       volume = {871},
       number = {2},
          eid = {249},
        pages = {249},
          doi = {10.3847/1538-4357/aafad7},
archivePrefix = {arXiv},
       eprint = {1812.11074},
 primaryClass = {astro-ph.GA},
       adsurl = {https://ui.adsabs.harvard.edu/abs/2019ApJ...871..249K}
}

@INPROCEEDINGS{paragi_2015,
       author = {{Paragi}, Z. and {Godfrey}, L. and {Reynolds}, C. and {Rioja}, M.~J. and {Deller}, A. and {Zhang}, B. and {Gurvits}, L. and {Bietenholz}, M. and {Szomoru}, A. and {Bignall}, H.~E. and {Boven}, P. and {Charlot}, P. and {Dodson}, R. and {Frey}, S. and {Garrett}, M.~A. and {Imai}, H. and {Lobanov}, A. and {Reid}, M.~J. and {Ros}, E. and {van Langevelde}, H.~J. and {Zensus}, A.~J. and {Zheng}, X.~W. and {Alberdi}, A. and {Agudo}, I. and {An}, T. and {Argo}, M. and {Beswick}, R. and {Biggs}, A. and {Brunthaler}, A. and {Campbell}, B. and {Cimo}, G. and {Colomer}, F. and {Corbel}, S. and {Conway}, J.~E. and {Cseh}, D. and {Deane}, R. and {Falcke}, H.~D.~E. and {Gawronski}, M. and {Gaylard}, M. and {Giovannini}, G. and {Giroletti}, M. and {Goddi}, C. and {Goedhart}, S. and {G{\'o}mez}, J.~L. and {Gunn}, A. and {Kharb}, P. and {Kloeckner}, H.~R. and {Koerding}, E. and {Kovalev}, Y. and {Kunert-Bajraszewska}, M. and {Lindqvist}, M. and {Lister}, M. and {Mantovani}, F. and {Marti-Vidal}, I. and {Mezcua}, M. and {McKean}, J. and {Middelberg}, E. and {Miller-Jones}, J.~C.~A. and {Moldon}, J. and {Muxlow}, T. and {O'Brien}, T. and {Perez-Torres}, M. and {Pogrebenko}, S.~V. and {Quick}, J. and {Rushton}, A. and {Schilizzi}, R. and {Smirnov}, O. and {Sohn}, B.~W. and {Surcis}, G. and {Taylor}, G.~B. and {Tingay}, S. and {Tudose}, V.~M. and {van der Horst}, A. and {van Leeuwen}, J. and {Venturi}, T. and {Vermeulen}, R. and {Vlemmings}, W.~H.~T. and {de Witt}, A. and {Wucknitz}, O. and {Yang}, J. and {Gab{\"a}nyi}, K. and {Jung}, T.},
        title = "{Very Long Baseline Interferometry with the SKA}",
    booktitle = {Advancing Astrophysics with the Square Kilometre Array (AASKA14)},
         year = 2015,
        month = apr,
          eid = {143},
        pages = {143},
          doi = {10.22323/1.215.0143},
archivePrefix = {arXiv},
       eprint = {1412.5971},
 primaryClass = {astro-ph.IM},
       adsurl = {https://ui.adsabs.harvard.edu/abs/2015aska.confE.143P}
}

@ARTICLE{Toubiana2024,
       author = {{Toubiana}, A. and {Sberna}, L. and {Volonteri}, M. and {Barausse}, E. and {Babak}, S. and {Enficiaud}, R. and {Izquierdo Villalba}, D. and {Gair}, J.~R. and {Greene}, J.~E. and {Quelquejay Leclere}, H.},
        title = "{Reconciling PTA and JWST and preparing for LISA with POMPOCO: a Parametrisation Of the Massive black hole POpulation for Comparison to Observations}",
      journal = {arXiv e-prints},
         year = 2024,
        month = oct,
          eid = {arXiv:2410.17916},
        pages = {arXiv:2410.17916},
          doi = {10.48550/arXiv.2410.17916},
archivePrefix = {arXiv},
       eprint = {2410.17916},
 primaryClass = {astro-ph.GA},
       adsurl = {https://ui.adsabs.harvard.edu/abs/2024arXiv241017916T}
}

@ARTICLE{Bonoli2025,
       author = {{Bonoli}, Silvia and {Izquierdo-Villalba}, David and {Spinoso}, Daniele and {Colpi}, Monica and {Sesana}, Alberto and {Polkas}, Markos and {Springel}, Volker},
        title = "{Constraints on the early growth of massive black holes from PTA and JWST with L-GalaxiesBH}",
      journal = {arXiv e-prints},
         year = 2025,
        month = sep,
          eid = {arXiv:2509.12325},
        pages = {arXiv:2509.12325},
          doi = {10.48550/arXiv.2509.12325},
archivePrefix = {arXiv},
       eprint = {2509.12325},
 primaryClass = {astro-ph.GA},
       adsurl = {https://ui.adsabs.harvard.edu/abs/2025arXiv250912325B}
}

@ARTICLE{Nandi24,
       author = {{Nandi}, Sumana and {Kharb}, Preeti and {Caproni}, Anderson and {Roy}, Rupak and {Sebastian}, Biny},
        title = "{A Relook at the Black Hole Binary Candidate J1328+2752 with VLBI}",
      journal = {\apj},
         year = 2024,
        month = apr,
       volume = {965},
       number = {1},
          eid = {9},
        pages = {9},
          doi = {10.3847/1538-4357/ad2c92},
       adsurl = {https://ui.adsabs.harvard.edu/abs/2024ApJ...965....9N}
}

@ARTICLE{Vaughan2016,
       author = {{Vaughan}, S. and {Uttley}, P. and {Markowitz}, A.~G. and {Huppenkothen}, D. and {Middleton}, M.~J. and {Alston}, W.~N. and {Scargle}, J.~D. and {Farr}, W.~M.},
        title = "{False periodicities in quasar time-domain surveys}",
      journal = {\mnras},
         year = 2016,
        month = sep,
       volume = {461},
       number = {3},
        pages = {3145-3152},
          doi = {10.1093/mnras/stw1412},
archivePrefix = {arXiv},
       eprint = {1606.02620},
 primaryClass = {astro-ph.IM},
       adsurl = {https://ui.adsabs.harvard.edu/abs/2016MNRAS.461.3145V}
}

@ARTICLE{ElBadry2025,
       author = {{El-Badry}, Kareem and {Hogg}, David W. and {Rix}, Hans-Walter},
        title = "{Active galactic nuclei do not exhibit stably periodic brightness variations}",
      journal = {arXiv e-prints},
         year = 2025,
        month = sep,
          eid = {arXiv:2509.10601},
        pages = {arXiv:2509.10601},
          doi = {10.48550/arXiv.2509.10601},
archivePrefix = {arXiv},
       eprint = {2509.10601},
 primaryClass = {astro-ph.GA},
       adsurl = {https://ui.adsabs.harvard.edu/abs/2025arXiv250910601E}
}

@ARTICLE{condon_1980,
       author = {{Condon}, J.~J. and {O'Dell}, S.~L. and {Puschell}, J.~J. and {Stein}, W.~A.},
        title = "{Radio emission from radio-quiet quasars}",
      journal = {\nat},
         year = 1980,
        month = jan,
       volume = {283},
       number = {5745},
        pages = {357-358},
          doi = {10.1038/283357a0},
       adsurl = {https://ui.adsabs.harvard.edu/abs/1980Natur.283..357C}
}

@ARTICLE{hale_2025,
       author = {{Hale}, C.~L. and {Heywood}, I. and {Jarvis}, M.~J. and {Whittam}, I.~H. and {Best}, P.~N. and {An}, Fangxia and {Bowler}, R.~A.~A. and {Harrison}, I. and {Matthews}, A. and {Smith}, D.~J.~B. and {Taylor}, A.~R. and {Vaccari}, M.},
        title = "{MIGHTEE: the continuum survey Data Release 1}",
      journal = {\mnras},
         year = 2025,
        month = jan,
       volume = {536},
       number = {3},
        pages = {2187-2211},
          doi = {10.1093/mnras/stae2528},
archivePrefix = {arXiv},
       eprint = {2411.04958},
 primaryClass = {astro-ph.GA},
       adsurl = {https://ui.adsabs.harvard.edu/abs/2025MNRAS.536.2187H}
}

@ARTICLE{jaffe_2003,
       author = {{Jaffe}, A.~H. and {Backer}, D.~C.},
        title = "{Gravitational Waves Probe the Coalescence Rate of Massive Black Hole Binaries}",
      journal = {\apj},
         year = 2003,
        month = feb,
       volume = {583},
       number = {2},
        pages = {616-631},
          doi = {10.1086/345443},
archivePrefix = {arXiv},
       eprint = {astro-ph/0210148},
 primaryClass = {astro-ph},
       adsurl = {https://ui.adsabs.harvard.edu/abs/2003ApJ...583..616J}
}

@ARTICLE{begelman_1980,
       author = {{Begelman}, M.~C. and {Blandford}, R.~D. and {Rees}, M.~J.},
        title = "{Massive black hole binaries in active galactic nuclei}",
      journal = {\nat},
         year = 1980,
        month = sep,
       volume = {287},
       number = {5780},
        pages = {307-309},
          doi = {10.1038/287307a0},
       adsurl = {https://ui.adsabs.harvard.edu/abs/1980Natur.287..307B}
}

@ARTICLE{kormendy_1995,
       author = {{Kormendy}, John and {Richstone}, Douglas},
        title = "{Inward Bound---The Search For Supermassive Black Holes In Galactic Nuclei}",
      journal = {\araa},
         year = 1995,
        month = jan,
       volume = {33},
        pages = {581},
          doi = {10.1146/annurev.aa.33.090195.003053},
       adsurl = {https://ui.adsabs.harvard.edu/abs/1995ARA&A..33..581K}
}

@ARTICLE{Iwasawa2011,
       author = {{Iwasawa}, K. and {Sanders}, D.~B. and {Teng}, S.~H. and {U}, Vivian and {Armus}, L. and {Evans}, A.~S. and {Howell}, J.~H. and {Komossa}, S. and {Mazzarella}, J.~M. and {Petric}, A.~O. and {Surace}, J.~A. and {Vavilkin}, T. and {Veilleux}, S. and {Trentham}, N.},
        title = "{C-GOALS: Chandra observations of a complete sample of luminous infrared galaxies from the IRAS Revised Bright Galaxy Survey}",
      journal = {\aap},
         year = 2011,
        month = may,
       volume = {529},
          eid = {A106},
        pages = {A106},
          doi = {10.1051/0004-6361/201015264},
archivePrefix = {arXiv},
       eprint = {1103.2755},
 primaryClass = {astro-ph.CO},
       adsurl = {https://ui.adsabs.harvard.edu/abs/2011A&A...529A.106I}
}

@ARTICLE{Sanders1996,
       author = {{Sanders}, D.~B. and {Mirabel}, I.~F.},
        title = "{Luminous Infrared Galaxies}",
      journal = {\araa},
         year = 1996,
        month = jan,
       volume = {34},
        pages = {749},
          doi = {10.1146/annurev.astro.34.1.749},
       adsurl = {https://ui.adsabs.harvard.edu/abs/1996ARA&A..34..749S}
}

@ARTICLE{Liu2014,
       author = {{Liu}, F.~K. and {Li}, Shuo and {Komossa}, S.},
        title = "{A Milliparsec Supermassive Black Hole Binary Candidate in the Galaxy SDSS J120136.02+300305.5}",
      journal = {\apj},
         year = 2014,
        month = may,
       volume = {786},
       number = {2},
          eid = {103},
        pages = {103},
          doi = {10.1088/0004-637X/786/2/103},
archivePrefix = {arXiv},
       eprint = {1404.4933},
 primaryClass = {astro-ph.HE},
       adsurl = {https://ui.adsabs.harvard.edu/abs/2014ApJ...786..103L}
}

@INPROCEEDINGS{Donnarumma2015,
       author = {{Donnarumma}, I. and {Rossi}, E.~M. and {Fender}, R. and {Komossa}, S. and {Paragi}, Z. and {Van Velzen}, S. and {Prandoni}, I.},
        title = "{SKA as a powerful hunter of jetted Tidal Disruption Events}",
    booktitle = {Advancing Astrophysics with the Square Kilometre Array (AASKA14)},
         year = 2015,
        month = apr,
          eid = {54},
        pages = {54},
          doi = {10.22323/1.215.0054},
archivePrefix = {arXiv},
       eprint = {1501.04640},
 primaryClass = {astro-ph.HE},
       adsurl = {https://ui.adsabs.harvard.edu/abs/2015aska.confE..54D}
}

@ARTICLE{Phinney2005,
       author = {{Milosavljevi{\'c}}, Milo{\v{s}} and {Phinney}, E.~S.},
        title = "{The Afterglow of Massive Black Hole Coalescence}",
      journal = {\apjl},
         year = 2005,
        month = apr,
       volume = {622},
       number = {2},
        pages = {L93-L96},
          doi = {10.1086/429618},
archivePrefix = {arXiv},
       eprint = {astro-ph/0410343},
 primaryClass = {astro-ph},
       adsurl = {https://ui.adsabs.harvard.edu/abs/2005ApJ...622L..93M}
}

@INPROCEEDINGS{Jenet2005,
       author = {{Jenet}, F.~A. and {Lommen}, A. and {Larson}, S.~L. and {Wen}, L.},
        title = "{Constraining the Properties of the Proposed Super-Massive Black Hole System in 3C66B: Limits from Pulsar Timing}",
    booktitle = {Binary Radio Pulsars},
         year = 2005,
       editor = {{Rasio}, Fred A. and {Stairs}, Ingrid H.},
       series = {Astronomical Society of the Pacific Conference Series},
       volume = {328},
        month = jul,
        pages = {399},
       adsurl = {https://ui.adsabs.harvard.edu/abs/2005ASPC..328..399J}
}

@ARTICLE{Sudou2003,
       author = {{Sudou}, Hiroshi and {Iguchi}, Satoru and {Murata}, Yasuhiro and {Taniguchi}, Yoshiaki},
        title = "{Orbital Motion in the Radio Galaxy 3C 66B: Evidence for a Supermassive Black Hole Binary}",
      journal = {Science},
         year = 2003,
        month = may,
       volume = {300},
       number = {5623},
        pages = {1263-1265},
          doi = {10.1126/science.1082817},
archivePrefix = {arXiv},
       eprint = {astro-ph/0306103},
 primaryClass = {astro-ph},
       adsurl = {https://ui.adsabs.harvard.edu/abs/2003Sci...300.1263S}
}

@ARTICLE{Ivezic2019,
       author = {{Ivezi{\'c}}, {\v{Z}}eljko and {Kahn}, Steven M. and {Tyson}, J. Anthony and {Abel}, Bob and {Acosta}, Emily and {Allsman}, Robyn and {Alonso}, David and {AlSayyad}, Yusra and {Anderson}, Scott F. and {Andrew}, John and {Angel}, James Roger P. and {Angeli}, George Z. and {Ansari}, Reza and {Antilogus}, Pierre and {Araujo}, Constanza and {Armstrong}, Robert and {Arndt}, Kirk T. and {Astier}, Pierre and {Aubourg}, {\'E}ric and {Auza}, Nicole and {Axelrod}, Tim S. and {Bard}, Deborah J. and {Barr}, Jeff D. and {Barrau}, Aurelian and {Bartlett}, James G. and {Bauer}, Amanda E. and {Bauman}, Brian J. and {Baumont}, Sylvain and {Bechtol}, Ellen and {Bechtol}, Keith and {Becker}, Andrew C. and {Becla}, Jacek and {Beldica}, Cristina and {Bellavia}, Steve and {Bianco}, Federica B. and {Biswas}, Rahul and {Blanc}, Guillaume and {Blazek}, Jonathan and {Blandford}, Roger D. and {Bloom}, Josh S. and {Bogart}, Joanne and {Bond}, Tim W. and {Booth}, Michael T. and {Borgland}, Anders W. and {Borne}, Kirk and {Bosch}, James F. and {Boutigny}, Dominique and {Brackett}, Craig A. and {Bradshaw}, Andrew and {Brandt}, William Nielsen and {Brown}, Michael E. and {Bullock}, James S. and {Burchat}, Patricia and {Burke}, David L. and {Cagnoli}, Gianpietro and {Calabrese}, Daniel and {Callahan}, Shawn and {Callen}, Alice L. and {Carlin}, Jeffrey L. and {Carlson}, Erin L. and {Chandrasekharan}, Srinivasan and {Charles-Emerson}, Glenaver and {Chesley}, Steve and {Cheu}, Elliott C. and {Chiang}, Hsin-Fang and {Chiang}, James and {Chirino}, Carol and {Chow}, Derek and {Ciardi}, David R. and {Claver}, Charles F. and {Cohen-Tanugi}, Johann and {Cockrum}, Joseph J. and {Coles}, Rebecca and {Connolly}, Andrew J. and {Cook}, Kem H. and {Cooray}, Asantha and {Covey}, Kevin R. and {Cribbs}, Chris and {Cui}, Wei and {Cutri}, Roc and {Daly}, Philip N. and {Daniel}, Scott F. and {Daruich}, Felipe and {Daubard}, Guillaume and {Daues}, Greg and {Dawson}, William and {Delgado}, Francisco and {Dellapenna}, Alfred and {de Peyster}, Robert and {de Val-Borro}, Miguel and {Digel}, Seth W. and {Doherty}, Peter and {Dubois}, Richard and {Dubois-Felsmann}, Gregory P. and {Durech}, Josef and {Economou}, Frossie and {Eifler}, Tim and {Eracleous}, Michael and {Emmons}, Benjamin L. and {Fausti Neto}, Angelo and {Ferguson}, Henry and {Figueroa}, Enrique and {Fisher-Levine}, Merlin and {Focke}, Warren and {Foss}, Michael D. and {Frank}, James and {Freemon}, Michael D. and {Gangler}, Emmanuel and {Gawiser}, Eric and {Geary}, John C. and {Gee}, Perry and {Geha}, Marla and {Gessner}, Charles J.~B. and {Gibson}, Robert R. and {Gilmore}, D. Kirk and {Glanzman}, Thomas and {Glick}, William and {Goldina}, Tatiana and {Goldstein}, Daniel A. and {Goodenow}, Iain and {Graham}, Melissa L. and {Gressler}, William J. and {Gris}, Philippe and {Guy}, Leanne P. and {Guyonnet}, Augustin and {Haller}, Gunther and {Harris}, Ron and {Hascall}, Patrick A. and {Haupt}, Justine and {Hernandez}, Fabio and {Herrmann}, Sven and {Hileman}, Edward and {Hoblitt}, Joshua and {Hodgson}, John A. and {Hogan}, Craig and {Howard}, James D. and {Huang}, Dajun and {Huffer}, Michael E. and {Ingraham}, Patrick and {Innes}, Walter R. and {Jacoby}, Suzanne H. and {Jain}, Bhuvnesh and {Jammes}, Fabrice and {Jee}, M. James and {Jenness}, Tim and {Jernigan}, Garrett and {Jevremovi{\'c}}, Darko and {Johns}, Kenneth and {Johnson}, Anthony S. and {Johnson}, Margaret W.~G. and {Jones}, R. Lynne and {Juramy-Gilles}, Claire and {Juri{\'c}}, Mario and {Kalirai}, Jason S. and {Kallivayalil}, Nitya J. and {Kalmbach}, Bryce and {Kantor}, Jeffrey P. and {Karst}, Pierre and {Kasliwal}, Mansi M. and {Kelly}, Heather and {Kessler}, Richard and {Kinnison}, Veronica and {Kirkby}, David and {Knox}, Lloyd and {Kotov}, Ivan V. and {Krabbendam}, Victor L. and {Krughoff}, K. Simon and {Kub{\'a}nek}, Petr and {Kuczewski}, John and {Kulkarni}, Shri and {Ku}, John and {Kurita}, Nadine R. and {Lage}, Craig S. and {Lambert}, Ron and {Lange}, Travis and {Langton}, J. Brian and {Le Guillou}, Laurent and {Levine}, Deborah and {Liang}, Ming and {Lim}, Kian-Tat and {Lintott}, Chris J. and {Long}, Kevin E. and {Lopez}, Margaux and {Lotz}, Paul J. and {Lupton}, Robert H. and {Lust}, Nate B. and {MacArthur}, Lauren A. and {Mahabal}, Ashish and {Mandelbaum}, Rachel and {Markiewicz}, Thomas W. and {Marsh}, Darren S. and {Marshall}, Philip J. and {Marshall}, Stuart and {May}, Morgan and {McKercher}, Robert and {McQueen}, Michelle and {Meyers}, Joshua and {Migliore}, Myriam and {Miller}, Michelle and {Mills}, David J.},
        title = "{LSST: From Science Drivers to Reference Design and Anticipated Data Products}",
      journal = {\apj},
         year = 2019,
        month = mar,
       volume = {873},
       number = {2},
          eid = {111},
        pages = {111},
          doi = {10.3847/1538-4357/ab042c},
archivePrefix = {arXiv},
       eprint = {0805.2366},
 primaryClass = {astro-ph},
       adsurl = {https://ui.adsabs.harvard.edu/abs/2019ApJ...873..111I}
}

@ARTICLE{Duan2025,
       author = {{Duan}, Qiao and {Conselice}, Christopher J. and {Li}, Qiong and {Austin}, Duncan and {Harvey}, Thomas and {Adams}, Nathan J. and {Duncan}, Kenneth J. and {Trussler}, James and {Ferreira}, Leonardo and {Westcott}, Lewi and {Harris}, Honor and {Windhorst}, Rogier A. and {Holwerda}, Benne W. and {Broadhurst}, Thomas J. and {Coe}, Dan and {Cohen}, Seth H. and {Du}, Xiaojing and {Driver}, Simon P. and {Frye}, Brenda and {Grogin}, Norman A. and {Hathi}, Nimish P. and {Jansen}, Rolf A. and {Koekemoer}, Anton M. and {Marshall}, Madeline A. and {Nonino}, Mario and {Ortiz}, III, Rafael and {Pirzkal}, Nor and {Robotham}, Aaron and {Ryan}, Russell E. and {Summers}, Jake and {D'Silva}, Jordan C.~J. and {Willmer}, Christopher N.~A. and {Yan}, Haojing},
        title = "{Galaxy mergers in the epoch of reionization {\textendash} I. A JWST study of pair fractions, merger rates, and stellar mass accretion rates at z = 4.5{\textendash}11.5}",
      journal = {\mnras},
         year = 2025,
        month = jun,
       volume = {540},
       number = {1},
        pages = {774-805},
          doi = {10.1093/mnras/staf638},
archivePrefix = {arXiv},
       eprint = {2407.09472},
 primaryClass = {astro-ph.GA},
       adsurl = {https://ui.adsabs.harvard.edu/abs/2025MNRAS.540..774D}
}

@ARTICLE{LiuChen2007,
       author = {{Liu}, F.~K. and {Chen}, X.},
        title = "{Evolution of Supermassive Black Hole Binaries and Acceleration of Jet Precession in Galactic Nuclei}",
      journal = {\apj},
         year = 2007,
        month = dec,
       volume = {671},
       number = {2},
        pages = {1272-1283},
          doi = {10.1086/522910},
archivePrefix = {arXiv},
       eprint = {0705.1077},
 primaryClass = {astro-ph},
       adsurl = {https://ui.adsabs.harvard.edu/abs/2007ApJ...671.1272L}
}

@ARTICLE{Graham2015a,
       author = {{Graham}, Matthew J. and {Djorgovski}, S.~G. and {Stern}, Daniel and {Glikman}, Eilat and {Drake}, Andrew J. and {Mahabal}, Ashish A. and {Donalek}, Ciro and {Larson}, Steve and {Christensen}, Eric},
        title = "{A possible close supermassive black-hole binary in a quasar with optical periodicity}",
      journal = {\nat},
         year = 2015,
        month = feb,
       volume = {518},
       number = {7537},
        pages = {74-76},
          doi = {10.1038/nature14143},
archivePrefix = {arXiv},
       eprint = {1501.01375},
 primaryClass = {astro-ph.GA},
       adsurl = {https://ui.adsabs.harvard.edu/abs/2015Natur.518...74G}
}

@ARTICLE{Rodriguez2006,
       author = {{Rodriguez}, C. and {Taylor}, G.~B. and {Zavala}, R.~T. and {Peck}, A.~B. and {Pollack}, L.~K. and {Romani}, R.~W.},
        title = "{A Compact Supermassive Binary Black Hole System}",
      journal = {\apj},
         year = 2006,
        month = jul,
       volume = {646},
       number = {1},
        pages = {49-60},
          doi = {10.1086/504825},
archivePrefix = {arXiv},
       eprint = {astro-ph/0604042},
 primaryClass = {astro-ph},
       adsurl = {https://ui.adsabs.harvard.edu/abs/2006ApJ...646...49R}
}

@ARTICLE{Panessa2013,
       author = {{Panessa}, Francesca and {Giroletti}, Marcello},
        title = "{Sub-parsec radio cores in nearby Seyfert galaxies}",
      journal = {\mnras},
         year = 2013,
        month = jun,
       volume = {432},
       number = {2},
        pages = {1138-1143},
          doi = {10.1093/mnras/stt547},
archivePrefix = {arXiv},
       eprint = {1304.0794},
 primaryClass = {astro-ph.HE},
       adsurl = {https://ui.adsabs.harvard.edu/abs/2013MNRAS.432.1138P}
}

@ARTICLE{Kharb2021,
       author = {{Kharb}, P. and {Subramanian}, S. and {Das}, M. and {Vaddi}, S. and {Paragi}, Z.},
        title = "{The Nature of Jets in Double-peaked Emission-line AGN in the KISSR Sample}",
      journal = {\apj},
         year = 2021,
        month = oct,
       volume = {919},
       number = {2},
          eid = {108},
        pages = {108},
          doi = {10.3847/1538-4357/ac0c82},
archivePrefix = {arXiv},
       eprint = {2106.09304},
 primaryClass = {astro-ph.GA},
       adsurl = {https://ui.adsabs.harvard.edu/abs/2021ApJ...919..108K}
}

@ARTICLE{Giroletti2009,
       author = {{Giroletti}, Marcello and {Panessa}, Francesca},
        title = "{The Faintest Seyfert Radio Cores Revealed by VLBI}",
      journal = {\apjl},
         year = 2009,
        month = dec,
       volume = {706},
       number = {2},
        pages = {L260-L264},
          doi = {10.1088/0004-637X/706/2/L260},
archivePrefix = {arXiv},
       eprint = {0910.5821},
 primaryClass = {astro-ph.CO},
       adsurl = {https://ui.adsabs.harvard.edu/abs/2009ApJ...706L.260G}
}

@misc{braun2019anticipatedperformancesquarekilometre,
      title={Anticipated Performance of the Square Kilometre Array -- Phase 1 (SKA1)}, 
      author={Robert Braun and Anna Bonaldi and Tyler Bourke and Evan Keane and Jeff Wagg},
      year={2019},
      eprint={1912.12699},
      archivePrefix={arXiv},
      primaryClass={astro-ph.IM},
      url={https://arxiv.org/abs/1912.12699}, 
}

@ARTICLE{hygate_2023,
       author = {{Hygate}, A.~P.~S. and {Hodge}, J.~A. and {da Cunha}, E. and {Rybak}, M. and {Schouws}, S. and {Inami}, H. and {Stefanon}, M. and {Graziani}, L. and {Schneider}, R. and {Dayal}, P. and {Bouwens}, R.~J. and {Smit}, R. and {Bowler}, R.~A.~A. and {Endsley}, R. and {Gonzalez}, V. and {Oesch}, P.~A. and {Stark}, D.~P. and {Algera}, H.~S.~B. and {Aravena}, M. and {Barrufet}, L. and {Ferrara}, A. and {Fudamoto}, Y. and {Hilhorst}, J.~H.~A. and {De Looze}, I. and {Nanayakkara}, T. and {Pallottini}, A. and {Riechers}, D.~A. and {Sommovigo}, L. and {Topping}, M.~W. and {van der Werf}, P.},
        title = "{The ALMA REBELS Survey: discovery of a massive, highly star-forming, and morphologically complex ULIRG at z = 7.31}",
      journal = {\mnras},
         year = 2023,
        month = sep,
       volume = {524},
       number = {2},
        pages = {1775-1795},
          doi = {10.1093/mnras/stad1212},
archivePrefix = {arXiv},
       eprint = {2304.09206},
 primaryClass = {astro-ph.GA},
       adsurl = {https://ui.adsabs.harvard.edu/abs/2023MNRAS.524.1775H}
}

@ARTICLE{sesana_2021,
       author = {{Sesana}, Alberto},
        title = "{Black hole science with the Laser Interferometer Space Antenna}",
      journal = {Frontiers in Astronomy and Space Sciences},
         year = 2021,
        month = feb,
       volume = {8},
          eid = {7},
        pages = {7},
          doi = {10.3389/fspas.2021.601646},
archivePrefix = {arXiv},
       eprint = {2105.11518},
 primaryClass = {astro-ph.CO},
       adsurl = {https://ui.adsabs.harvard.edu/abs/2021FrASS...8....7S}
}

@ARTICLE{tang_2025,
       author = {{Tang}, Shenli and {Silverman}, John D. and {Liu}, Zhaoxuan and {Banerji}, Manda and {Suzuki}, Tomoko and {Fujimoto}, Seiji and {Goulding}, Andy and {Imanishi}, Masatoshi and {Kawaguchi}, Toshihiro and {Bottrell}, Connor and {Hartwig}, Tilman and {Jahnke}, Knud and {Onoue}, Masafusa and {Schramm}, Malte and {Ueda}, Yoshihiro},
        title = "{ALMA observations of dual quasars: evidence of rich and diverse molecular gas environments}",
      journal = {\mnras},
         year = 2025,
        month = apr,
       volume = {538},
       number = {4},
        pages = {3001-3022},
          doi = {10.1093/mnras/staf416},
archivePrefix = {arXiv},
       eprint = {2407.09399},
 primaryClass = {astro-ph.GA},
       adsurl = {https://ui.adsabs.harvard.edu/abs/2025MNRAS.538.3001T}
}

@ARTICLE{li_2024,
       author = {{Li}, Yingjie and {Xu}, Ye and {Li}, Jingjing and {Bian}, Shuaibo and {Lin}, Zehao and {Hao}, Chaojie and {Liu}, Dejian},
        title = "{VLBI with SKA: Possible Arrays and Astrometric Science}",
      journal = {Research in Astronomy and Astrophysics},
         year = 2024,
        month = jul,
       volume = {24},
       number = {7},
          eid = {072001},
        pages = {072001},
          doi = {10.1088/1674-4527/ad420c},
archivePrefix = {arXiv},
       eprint = {2404.14663},
 primaryClass = {astro-ph.GA},
       adsurl = {https://ui.adsabs.harvard.edu/abs/2024RAA....24g2001L}
}

@ARTICLE{deane2014nature,
       author = {{Deane}, R.~P. and {Paragi}, Z. and {Jarvis}, M.~J. and {Coriat}, M. and {Bernardi}, G. and {Fender}, R.~P. and {Frey}, S. and {Heywood}, I. and {Kl{\"o}ckner}, H. -R. and {Grainge}, K. and {Rumsey}, C.},
        title = "{A close-pair binary in a distant triple supermassive black hole system}",
      journal = {\nat},
         year = 2014,
        month = jul,
       volume = {511},
       number = {7507},
        pages = {57-60},
          doi = {10.1038/nature13454},
archivePrefix = {arXiv},
       eprint = {1406.6365},
 primaryClass = {astro-ph.GA},
       adsurl = {https://ui.adsabs.harvard.edu/abs/2014Natur.511...57D}
}

@ARTICLE{spingola2019,
       author = {{Spingola}, C. and {McKean}, J.~P. and {Massari}, D. and {Koopmans}, L.~V.~E.},
        title = "{Proper motion in lensed radio jets at redshift 3: A possible dual super-massive black hole system in the early Universe}",
      journal = {\aap},
         year = 2019,
        month = oct,
       volume = {630},
          eid = {A108},
        pages = {A108},
          doi = {10.1051/0004-6361/201935427},
archivePrefix = {arXiv},
       eprint = {1908.11756},
 primaryClass = {astro-ph.GA},
       adsurl = {https://ui.adsabs.harvard.edu/abs/2019A&A...630A.108S}
}

@ARTICLE{derosa_2019,
       author = {{De Rosa}, Alessandra and {Vignali}, Cristian and {Bogdanovi{\'c}}, Tamara and {Capelo}, Pedro R. and {Charisi}, Maria and {Dotti}, Massimo and {Husemann}, Bernd and {Lusso}, Elisabeta and {Mayer}, Lucio and {Paragi}, Zsolt and {Runnoe}, Jessie and {Sesana}, Alberto and {Steinborn}, Lisa and {Bianchi}, Stefano and {Colpi}, Monica and {del Valle}, Luciano and {Frey}, S{\'a}ndor and {Gab{\'a}nyi}, Krisztina {\'E}. and {Giustini}, Margherita and {Guainazzi}, Matteo and {Haiman}, Zoltan and {Herrera Ruiz}, Noelia and {Herrero-Illana}, Rub{\'e}n and {Iwasawa}, Kazushi and {Komossa}, S. and {Lena}, Davide and {Loiseau}, Nora and {Perez-Torres}, Miguel and {Piconcelli}, Enrico and {Volonteri}, Marta},
        title = "{The quest for dual and binary supermassive black holes: A multi-messenger view}",
      journal = {\nar},
         year = 2019,
        month = dec,
       volume = {86},
          eid = {101525},
        pages = {101525},
          doi = {10.1016/j.newar.2020.101525},
archivePrefix = {arXiv},
       eprint = {2001.06293},
 primaryClass = {astro-ph.GA},
       adsurl = {https://ui.adsabs.harvard.edu/abs/2019NewAR..8601525D}
}

@ARTICLE{KH_2013,
       author = {{Kormendy}, John and {Ho}, Luis C.},
        title = "{Coevolution (Or Not) of Supermassive Black Holes and Host Galaxies}",
      journal = {\araa},
         year = 2013,
        month = aug,
       volume = {51},
       number = {1},
        pages = {511-653},
          doi = {10.1146/annurev-astro-082708-101811},
archivePrefix = {arXiv},
       eprint = {1304.7762},
 primaryClass = {astro-ph.CO},
       adsurl = {https://ui.adsabs.harvard.edu/abs/2013ARA&A..51..511K}
}

@ARTICLE{volonteri_2003,
       author = {{Volonteri}, Marta and {Madau}, Piero and {Haardt}, Francesco},
        title = "{The Formation of Galaxy Stellar Cores by the Hierarchical Merging of Supermassive Black Holes}",
      journal = {\apj},
         year = 2003,
        month = aug,
       volume = {593},
       number = {2},
        pages = {661-666},
          doi = {10.1086/376722},
archivePrefix = {arXiv},
       eprint = {astro-ph/0304389},
 primaryClass = {astro-ph},
       adsurl = {https://ui.adsabs.harvard.edu/abs/2003ApJ...593..661V}
}

@ARTICLE{mayer_2007,
       author = {{Mayer}, L. and {Kazantzidis}, S. and {Madau}, P. and {Colpi}, M. and {Quinn}, T. and {Wadsley}, J.},
        title = "{Rapid Formation of Supermassive Black Hole Binaries in Galaxy Mergers with Gas}",
      journal = {Science},
         year = 2007,
        month = jun,
       volume = {316},
       number = {5833},
        pages = {1874},
          doi = {10.1126/science.1141858},
archivePrefix = {arXiv},
       eprint = {0706.1562},
 primaryClass = {astro-ph},
       adsurl = {https://ui.adsabs.harvard.edu/abs/2007Sci...316.1874M}
}

@ARTICLE{tremmel_2017,
       author = {{Tremmel}, M. and {Karcher}, M. and {Governato}, F. and {Volonteri}, M. and {Quinn}, T.~R. and {Pontzen}, A. and {Anderson}, L. and {Bellovary}, J.},
        title = "{The Romulus cosmological simulations: a physical approach to the formation, dynamics and accretion models of SMBHs}",
      journal = {\mnras},
         year = 2017,
        month = sep,
       volume = {470},
       number = {1},
        pages = {1121-1139},
          doi = {10.1093/mnras/stx1160},
archivePrefix = {arXiv},
       eprint = {1607.02151},
 primaryClass = {astro-ph.GA},
       adsurl = {https://ui.adsabs.harvard.edu/abs/2017MNRAS.470.1121T}
}

@ARTICLE{volonteri_2022,
       author = {{Volonteri}, Marta and {Pfister}, Hugo and {Beckmann}, Ricarda and {Dotti}, Massimo and {Dubois}, Yohan and {Massonneau}, Warren and {Musoke}, Gibwa and {Tremmel}, Michael},
        title = "{Dual AGN in the Horizon-AGN simulation and their link to galaxy and massive black hole mergers, with an excursus on multiple AGN}",
      journal = {\mnras},
         year = 2022,
        month = jul,
       volume = {514},
       number = {1},
        pages = {640-656},
          doi = {10.1093/mnras/stac1217},
archivePrefix = {arXiv},
       eprint = {2112.07193},
 primaryClass = {astro-ph.GA},
       adsurl = {https://ui.adsabs.harvard.edu/abs/2022MNRAS.514..640V}
}

@ARTICLE{dotti_2007,
       author = {{Dotti}, M. and {Colpi}, M. and {Haardt}, F. and {Mayer}, L.},
        title = "{Supermassive black hole binaries in gaseous and stellar circumnuclear discs: orbital dynamics and gas accretion}",
      journal = {\mnras},
         year = 2007,
        month = aug,
       volume = {379},
       number = {3},
        pages = {956-962},
          doi = {10.1111/j.1365-2966.2007.12010.x},
archivePrefix = {arXiv},
       eprint = {astro-ph/0612505},
 primaryClass = {astro-ph},
       adsurl = {https://ui.adsabs.harvard.edu/abs/2007MNRAS.379..956D}
}

@ARTICLE{shen_2023,
       author = {{Shen}, Yue and {Hwang}, Hsiang-Chih and {Oguri}, Masamune and {Chen}, Nianyi and {Di Matteo}, Tiziana and {Ni}, Yueying and {Bird}, Simeon and {Zakamska}, Nadia and {Liu}, Xin and {Chen}, Yu-Ching and {Kratter}, Kaitlin M.},
        title = "{Statistics of Galactic-scale Quasar Pairs at Cosmic Noon}",
      journal = {\apj},
         year = 2023,
        month = jan,
       volume = {943},
       number = {1},
          eid = {38},
        pages = {38},
          doi = {10.3847/1538-4357/aca662},
archivePrefix = {arXiv},
       eprint = {2208.04979},
 primaryClass = {astro-ph.GA},
       adsurl = {https://ui.adsabs.harvard.edu/abs/2023ApJ...943...38S}
}

@ARTICLE{dimatteo_2023,
       author = {{Di Matteo}, Tiziana and {Angles-Alcazar}, Daniel and {Shankar}, Francesco},
        title = "{Massive black holes in galactic nuclei: Theory and Simulations}",
      journal = {arXiv e-prints},
         year = 2023,
        month = apr,
          eid = {arXiv:2304.11541},
        pages = {arXiv:2304.11541},
          doi = {10.48550/arXiv.2304.11541},
archivePrefix = {arXiv},
       eprint = {2304.11541},
 primaryClass = {astro-ph.HE},
       adsurl = {https://ui.adsabs.harvard.edu/abs/2023arXiv230411541D}
}

@ARTICLE{agazie_2023,
       author = {{Agazie}, Gabriella and {Alam}, Md Faisal and {Anumarlapudi}, Akash and {Archibald}, Anne M. and {Arzoumanian}, Zaven and {Baker}, Paul T. and {Blecha}, Laura and {Bonidie}, Victoria and {Brazier}, Adam and {Brook}, Paul R. and {Burke-Spolaor}, Sarah and {B{\'e}csy}, Bence and {Chapman}, Christopher and {Charisi}, Maria and {Chatterjee}, Shami and {Cohen}, Tyler and {Cordes}, James M. and {Cornish}, Neil J. and {Crawford}, Fronefield and {Cromartie}, H. Thankful and {Crowter}, Kathryn and {Decesar}, Megan E. and {Demorest}, Paul B. and {Dolch}, Timothy and {Drachler}, Brendan and {Ferrara}, Elizabeth C. and {Fiore}, William and {Fonseca}, Emmanuel and {Freedman}, Gabriel E. and {Garver-Daniels}, Nate and {Gentile}, Peter A. and {Glaser}, Joseph and {Good}, Deborah C. and {G{\"u}ltekin}, Kayhan and {Hazboun}, Jeffrey S. and {Jennings}, Ross J. and {Jessup}, Cody and {Johnson}, Aaron D. and {Jones}, Megan L. and {Kaiser}, Andrew R. and {Kaplan}, David L. and {Kelley}, Luke Zoltan and {Kerr}, Matthew and {Key}, Joey S. and {Kuske}, Anastasia and {Laal}, Nima and {Lam}, Michael T. and {Lamb}, William G. and {Lazio}, T. Joseph W. and {Lewandowska}, Natalia and {Lin}, Ye and {Liu}, Tingting and {Lorimer}, Duncan R. and {Luo}, Jing and {Lynch}, Ryan S. and {Ma}, Chung-Pei and {Madison}, Dustin R. and {Maraccini}, Kaleb and {McEwen}, Alexander and {McKee}, James W. and {McLaughlin}, Maura A. and {McMann}, Natasha and {Meyers}, Bradley W. and {Mingarelli}, Chiara M.~F. and {Mitridate}, Andrea and {Ng}, Cherry and {Nice}, David J. and {Ocker}, Stella Koch and {Olum}, Ken D. and {Panciu}, Elisa and {Pennucci}, Timothy T. and {Perera}, Benetge B.~P. and {Pol}, Nihan S. and {Radovan}, Henri A. and {Ransom}, Scott M. and {Ray}, Paul S. and {Romano}, Joseph D. and {Salo}, Laura and {Sardesai}, Shashwat C. and {Schmiedekamp}, Carl and {Schmiedekamp}, Ann and {Schmitz}, Kai and {Shapiro-Albert}, Brent J. and {Siemens}, Xavier and {Simon}, Joseph and {Siwek}, Magdalena S. and {Stairs}, Ingrid H. and {Stinebring}, Daniel R. and {Stovall}, Kevin and {Susobhanan}, Abhimanyu and {Swiggum}, Joseph K. and {Taylor}, Stephen R. and {Turner}, Jacob E. and {Unal}, Caner and {Vallisneri}, Michele and {Vigeland}, Sarah J. and {Wahl}, Haley M. and {Wang}, Qiaohong and {Witt}, Caitlin A. and {Young}, Olivia and {Nanograv Collaboration}},
        title = "{The NANOGrav 15 yr Data Set: Observations and Timing of 68 Millisecond Pulsars}",
      journal = {\apjl},
         year = 2023,
        month = jul,
       volume = {951},
       number = {1},
          eid = {L9},
        pages = {L9},
          doi = {10.3847/2041-8213/acda9a},
archivePrefix = {arXiv},
       eprint = {2306.16217},
 primaryClass = {astro-ph.HE},
       adsurl = {https://ui.adsabs.harvard.edu/abs/2023ApJ...951L...9A}
}

@ARTICLE{EPTA_collab_2023,
       author = {{EPTA Collaboration} and {InPTA Collaboration} and {Antoniadis}, J. and {Arumugam}, P. and {Arumugam}, S. and {Babak}, S. and {Bagchi}, M. and {Bak Nielsen}, A. -S. and {Bassa}, C.~G. and {Bathula}, A. and {Berthereau}, A. and {Bonetti}, M. and {Bortolas}, E. and {Brook}, P.~R. and {Burgay}, M. and {Caballero}, R.~N. and {Chalumeau}, A. and {Champion}, D.~J. and {Chanlaridis}, S. and {Chen}, S. and {Cognard}, I. and {Dandapat}, S. and {Deb}, D. and {Desai}, S. and {Desvignes}, G. and {Dhanda-Batra}, N. and {Dwivedi}, C. and {Falxa}, M. and {Ferdman}, R.~D. and {Franchini}, A. and {Gair}, J.~R. and {Goncharov}, B. and {Gopakumar}, A. and {Graikou}, E. and {Grie{\ss}meier}, J. -M. and {Guillemot}, L. and {Guo}, Y.~J. and {Gupta}, Y. and {Hisano}, S. and {Hu}, H. and {Iraci}, F. and {Izquierdo-Villalba}, D. and {Jang}, J. and {Jawor}, J. and {Janssen}, G.~H. and {Jessner}, A. and {Joshi}, B.~C. and {Kareem}, F. and {Karuppusamy}, R. and {Keane}, E.~F. and {Keith}, M.~J. and {Kharbanda}, D. and {Kikunaga}, T. and {Kolhe}, N. and {Kramer}, M. and {Krishnakumar}, M.~A. and {Lackeos}, K. and {Lee}, K.~J. and {Liu}, K. and {Liu}, Y. and {Lyne}, A.~G. and {McKee}, J.~W. and {Maan}, Y. and {Main}, R.~A. and {Mickaliger}, M.~B. and {Ni{\c{t}}u}, I.~C. and {Nobleson}, K. and {Paladi}, A.~K. and {Parthasarathy}, A. and {Perera}, B.~B.~P. and {Perrodin}, D. and {Petiteau}, A. and {Porayko}, N.~K. and {Possenti}, A. and {Prabu}, T. and {Quelquejay Leclere}, H. and {Rana}, P. and {Samajdar}, A. and {Sanidas}, S.~A. and {Sesana}, A. and {Shaifullah}, G. and {Singha}, J. and {Speri}, L. and {Spiewak}, R. and {Srivastava}, A. and {Stappers}, B.~W. and {Surnis}, M. and {Susarla}, S.~C. and {Susobhanan}, A. and {Takahashi}, K. and {Tarafdar}, P. and {Theureau}, G. and {Tiburzi}, C. and {van der Wateren}, E. and {Vecchio}, A. and {Venkatraman Krishnan}, V. and {Verbiest}, J.~P.~W. and {Wang}, J. and {Wang}, L. and {Wu}, Z.},
        title = "{The second data release from the European Pulsar Timing Array. III. Search for gravitational wave signals}",
      journal = {\aap},
         year = 2023,
        month = oct,
       volume = {678},
          eid = {A50},
        pages = {A50},
          doi = {10.1051/0004-6361/202346844},
archivePrefix = {arXiv},
       eprint = {2306.16214},
 primaryClass = {astro-ph.HE},
       adsurl = {https://ui.adsabs.harvard.edu/abs/2023A&A...678A..50E}
}

@ARTICLE{colpi_2019,
       author = {{Colpi}, Monica and {Holley-Bockelmann}, Kelly and {Bogdanovic}, Tamara and {Natarajan}, Priya and {Bellovary}, Jillian and {Sesana}, Alberto and {Tremmel}, Michael and {Schnittman}, Jeremy and {Comerford}, Julia and {Barausse}, Enrico and {Berti}, Emanuele and {Volonteri}, Marta and {Khan}, Fazeel and {McWilliams}, Sean and {Burke-Spolaor}, Sarah and {Hazboun}, Jeff and {Conklin}, John and {Mueller}, Guido and {Larson}, Shane},
        title = "{Astro2020 science white paper: The gravitational wave view of massive black holes}",
      journal = {arXiv e-prints},
         year = 2019,
        month = mar,
          eid = {arXiv:1903.06867},
        pages = {arXiv:1903.06867},
          doi = {10.48550/arXiv.1903.06867},
archivePrefix = {arXiv},
       eprint = {1903.06867},
 primaryClass = {astro-ph.GA},
       adsurl = {https://ui.adsabs.harvard.edu/abs/2019arXiv190306867C}
}

@ARTICLE{shen_2019,
       author = {{Shen}, Yue and {Hwang}, Hsiang-Chih and {Zakamska}, Nadia and {Liu}, Xin},
        title = "{Varstrometry for Off-nucleus and Dual Sub-Kpc AGN (VODKA): How Well Centered Are Low-z AGN?}",
      journal = {\apjl},
         year = 2019,
        month = nov,
       volume = {885},
       number = {1},
          eid = {L4},
        pages = {L4},
          doi = {10.3847/2041-8213/ab4b54},
archivePrefix = {arXiv},
       eprint = {1910.02969},
 primaryClass = {astro-ph.GA},
       adsurl = {https://ui.adsabs.harvard.edu/abs/2019ApJ...885L...4S}
}

@ARTICLE{mannucci_2022,
       author = {{Mannucci}, F. and {Pancino}, E. and {Belfiore}, F. and {Cicone}, C. and {Ciurlo}, A. and {Cresci}, G. and {Lusso}, E. and {Marasco}, A. and {Marconi}, A. and {Nardini}, E. and {Pinna}, E. and {Severgnini}, P. and {Saracco}, P. and {Tozzi}, G. and {Yeh}, S.},
        title = "{Unveiling the population of dual and lensed active galactic nuclei at sub-arcsec separations}",
      journal = {Nature Astronomy},
         year = 2022,
        month = aug,
       volume = {6},
        pages = {1185-1192},
          doi = {10.1038/s41550-022-01761-5},
archivePrefix = {arXiv},
       eprint = {2203.11234},
 primaryClass = {astro-ph.GA},
       adsurl = {https://ui.adsabs.harvard.edu/abs/2022NatAs...6.1185M}
}

@ARTICLE{mannucci_2023,
       author = {{Mannucci}, F. and {Scialpi}, M. and {Ciurlo}, A. and {Yeh}, S. and {Marconcini}, C. and {Tozzi}, G. and {Cresci}, G. and {Marconi}, A. and {Amiri}, A. and {Belfiore}, F. and {Carniani}, S. and {Cicone}, C. and {Nardini}, E. and {Pancino}, E. and {Rubinur}, K. and {Severgnini}, P. and {Ulivi}, L. and {Venturi}, G. and {Vignali}, C. and {Volonteri}, M. and {Pinna}, E. and {Rossi}, F. and {Puglisi}, A. and {Agapito}, G. and {Plantet}, C. and {Ghose}, E. and {Carbonaro}, L. and {Xompero}, M. and {Grani}, P. and {Esposito}, S. and {Power}, J. and {Guerra Ramon}, J.~C. and {Lefebvre}, M. and {Cavallaro}, A. and {Davies}, R. and {Riccardi}, A. and {Macintosh}, M. and {Taylor}, W. and {Dolci}, M. and {Baruffolo}, A. and {Feuchtgruber}, H. and {Kravchenko}, K. and {Rau}, C. and {Sturm}, E. and {Wiezorrek}, E. and {Dallilar}, Y. and {Kenworthy}, M.},
        title = "{GMP-selected dual and lensed AGNs: Selection function and classification based on near-IR colors and resolved spectra from VLT/ERIS, Keck/OSIRIS, and LBT/LUCI}",
      journal = {\aap},
         year = 2023,
        month = dec,
       volume = {680},
          eid = {A53},
        pages = {A53},
          doi = {10.1051/0004-6361/202346894},
archivePrefix = {arXiv},
       eprint = {2305.07396},
 primaryClass = {astro-ph.GA},
       adsurl = {https://ui.adsabs.harvard.edu/abs/2023A&A...680A..53M}
}

@ARTICLE{meier_2001,
       author = {{Meier}, D.~L.},
        title = "{The Association of Jet Production with Geometrically Thick Accretion Flows and Black Hole Rotation}",
      journal = {\apjl},
         year = 2001,
        month = feb,
       volume = {548},
       number = {1},
        pages = {L9-L12},
          doi = {10.1086/318921},
archivePrefix = {arXiv},
       eprint = {astro-ph/0010231},
 primaryClass = {astro-ph},
       adsurl = {https://ui.adsabs.harvard.edu/abs/2001ApJ...548L...9M}
}

@ARTICLE{puerto_sanchez_2025,
       author = {{Puerto-S{\'a}nchez}, Clara and {Habouzit}, M{\'e}lanie and {Volonteri}, Marta and {Ni}, Yueying and {Foord}, Adi and {Angl{\'e}s-Alc{\'a}zar}, Daniel and {Chen}, Nianyi and {Guetzoyan}, Paloma and {Dav{\'e}}, Romeel and {Di Matteo}, Tiziana and {Dubois}, Yohan and {Koss}, Michael and {Rosas-Guevara}, Yetli},
        title = "{Large-scale dual AGN in large-scale cosmological hydrodynamical simulations}",
      journal = {\mnras},
         year = 2025,
        month = jan,
       volume = {536},
       number = {3},
        pages = {3016-3040},
          doi = {10.1093/mnras/stae2763},
archivePrefix = {arXiv},
       eprint = {2411.15297},
 primaryClass = {astro-ph.CO},
       adsurl = {https://ui.adsabs.harvard.edu/abs/2025MNRAS.536.3016P}
}

@incollection{Shu01.2026.SKA, author = {Xinwen Shu and author2 and author3 and author4 and author5},title = {},year = {2026},publisher = {},note = {arXiv search: Report number AASKAII/Shu01},booktitle = {Advancing Astrophysics with the SKA -- II (AASKAII)}}

@incollection{Spingola01.2026.SKA, author = {Cristiana Spingola and author2 and author3 and author4 and author5},title = {},year = {2026},publisher = {},note = {arXiv search: Report number AASKAII/Spingola01},booktitle = {Advancing Astrophysics with the SKA -- II (AASKAII)}}

@incollection{Shannon01.2026.SKA, author = {Ryan M. Shannon and author2 and author3 and author4 and author5},title = {},year = {2026},publisher = {},note = {arXiv search: Report number AASKAII/Shannon01},booktitle = {Advancing Astrophysics with the SKA -- II (AASKAII)}}

@incollection{Kobayashi01.2026.SKA, author = {Hideyuki Kobayashi and author2 and author3 and author4 and author5},title = {},year = {2026},publisher = {},note = {arXiv search: Report number AASKAII/Kobayashi01},booktitle = {Advancing Astrophysics with the SKA -- II (AASKAII)}}

@incollection{Takahashi01.2026.SKA, author = {Keitaro Takahashi and author2 and author3 and author4 and author5},title = {},year = {2026},publisher = {},note = {arXiv search: Report number AASKAII/Takahashi01},booktitle = {Advancing Astrophysics with the SKA -- II (AASKAII)}}

@incollection{GemmaAnderson01.2026.SKA, author = {Gemma E. Anderson and author2 and author3 and author4 and author5},title = {},year = {2026},publisher = {},note = {arXiv search: Report number AASKAII/GemmaAnderson01},booktitle = {Advancing Astrophysics with the SKA -- II (AASKAII)}}

@ARTICLE{shen_2020,
       author = {{Shen}, Xuejian and {Hopkins}, Philip F. and {Faucher-Gigu{\`e}re}, Claude-Andr{\'e} and {Alexander}, D.~M. and {Richards}, Gordon T. and {Ross}, Nicholas P. and {Hickox}, R.~C.},
        title = "{The bolometric quasar luminosity function at z = 0-7}",
      journal = {\mnras},
         year = 2020,
        month = jan,
       volume = {495},
       number = {3},
        pages = {3252-3275},
          doi = {10.1093/mnras/staa1381},
archivePrefix = {arXiv},
       eprint = {2001.02696},
 primaryClass = {astro-ph.GA},
       adsurl = {https://ui.adsabs.harvard.edu/abs/2020MNRAS.495.3252S}
}

@ARTICLE{smolcic_2017,
       author = {{Smol{\v{c}}i{\'c}}, V. and {Delvecchio}, I. and {Zamorani}, G. and {Baran}, N. and {Novak}, M. and {Delhaize}, J. and {Schinnerer}, E. and {Berta}, S. and {Bondi}, M. and {Ciliegi}, P. and {Capak}, P. and {Civano}, F. and {Karim}, A. and {Le Fevre}, O. and {Ilbert}, O. and {Laigle}, C. and {Marchesi}, S. and {McCracken}, H.~J. and {Tasca}, L. and {Salvato}, M. and {Vardoulaki}, E.},
        title = "{The VLA-COSMOS 3 GHz Large Project: Multiwavelength counterparts and the composition of the faint radio population}",
      journal = {\aap},
         year = 2017,
        month = jun,
       volume = {602},
          eid = {A2},
        pages = {A2},
          doi = {10.1051/0004-6361/201630223},
archivePrefix = {arXiv},
       eprint = {1703.09719},
 primaryClass = {astro-ph.GA},
       adsurl = {https://ui.adsabs.harvard.edu/abs/2017A&A...602A...2S}
}

@ARTICLE{fu_2012,
       author = {{Fu}, Hai and {Yan}, Lin and {Myers}, Adam D. and {Stockton}, Alan and {Djorgovski}, S.~G. and {Aldering}, G. and {Rich}, Jeffrey A.},
        title = "{The Nature of Double-peaked [O III] Active Galactic Nuclei}",
      journal = {\apj},
         year = 2012,
        month = jan,
       volume = {745},
       number = {1},
          eid = {67},
        pages = {67},
          doi = {10.1088/0004-637X/745/1/67},
archivePrefix = {arXiv},
       eprint = {1107.3564},
 primaryClass = {astro-ph.CO},
       adsurl = {https://ui.adsabs.harvard.edu/abs/2012ApJ...745...67F}
}

@ARTICLE{sesana_2013,
       author = {{Sesana}, A.},
        title = "{Systematic investigation of the expected gravitational wave signal from  supermassive black hole binaries in the pulsar timing band.}",
      journal = {\mnras},
         year = 2013,
        month = jun,
       volume = {433},
        pages = {L1-L5},
          doi = {10.1093/mnrasl/slt034},
archivePrefix = {arXiv},
       eprint = {1211.5375},
 primaryClass = {astro-ph.CO},
       adsurl = {https://ui.adsabs.harvard.edu/abs/2013MNRAS.433L...1S}
}

@BOOK{merritt_2013,
       author = {{Merritt}, David},
        title = "{Dynamics and Evolution of Galactic Nuclei}",
         year = 2013,
       adsurl = {https://ui.adsabs.harvard.edu/abs/2013degn.book.....M}
}

@ARTICLE{merloni_2003,
       author = {{Merloni}, Andrea and {Heinz}, Sebastian and {di Matteo}, Tiziana},
        title = "{A Fundamental Plane of black hole activity}",
      journal = {\mnras},
         year = 2003,
        month = nov,
       volume = {345},
       number = {4},
        pages = {1057-1076},
          doi = {10.1046/j.1365-2966.2003.07017.x},
archivePrefix = {arXiv},
       eprint = {astro-ph/0305261},
 primaryClass = {astro-ph},
       adsurl = {https://ui.adsabs.harvard.edu/abs/2003MNRAS.345.1057M}
}

@ARTICLE{panessa_2015,
       author = {{Panessa}, F. and {Tarchi}, A. and {Castangia}, P. and {Maiorano}, E. and {Bassani}, L. and {Bicknell}, G. and {Bazzano}, A. and {Bird}, A.~J. and {Malizia}, A. and {Ubertini}, P.},
        title = "{The 1.4-GHz radio properties of hard X-ray-selected AGN}",
      journal = {\mnras},
         year = 2015,
        month = feb,
       volume = {447},
       number = {2},
        pages = {1289-1298},
          doi = {10.1093/mnras/stu2455},
archivePrefix = {arXiv},
       eprint = {1411.7829},
 primaryClass = {astro-ph.HE},
       adsurl = {https://ui.adsabs.harvard.edu/abs/2015MNRAS.447.1289P}
}

@ARTICLE{chen_2023_sim,
       author = {{Chen}, Nianyi and {Di Matteo}, Tiziana and {Ni}, Yueying and {Tremmel}, Michael and {DeGraf}, Colin and {Shen}, Yue and {Holgado}, A. Miguel and {Bird}, Simeon and {Croft}, Rupert and {Feng}, Yu},
        title = "{Properties and evolution of dual and offset AGN in the ASTRID simulation at z   2}",
      journal = {\mnras},
         year = 2023,
        month = jun,
       volume = {522},
       number = {2},
        pages = {1895-1913},
          doi = {10.1093/mnras/stad834},
archivePrefix = {arXiv},
       eprint = {2208.04970},
 primaryClass = {astro-ph.GA},
       adsurl = {https://ui.adsabs.harvard.edu/abs/2023MNRAS.522.1895C}
}

@ARTICLE{yue_2023,
       author = {{Yue}, Minghao and {Fan}, Xiaohui and {Yang}, Jinyi and {Wang}, Feige},
        title = "{A Survey for High-redshift Gravitationally Lensed Quasars and Close Quasar Pairs. I. The Discoveries of an Intermediately Lensed Quasar and a Kiloparsec-scale Quasar Pair at z   5}",
      journal = {\aj},
         year = 2023,
        month = may,
       volume = {165},
       number = {5},
          eid = {191},
        pages = {191},
          doi = {10.3847/1538-3881/acc2be},
archivePrefix = {arXiv},
       eprint = {2303.04357},
 primaryClass = {astro-ph.GA},
       adsurl = {https://ui.adsabs.harvard.edu/abs/2023AJ....165..191Y}
}

@ARTICLE{perna_2025,
       author = {{Perna}, Michele and {Arribas}, Santiago and {Lamperti}, Isabella and {Circosta}, Chiara and {Bertola}, Elena and {P{\'e}rez-Gonz{\'a}lez}, Pablo G. and {D'Eugenio}, Francesco and {{\"U}bler}, Hannah and {Cresci}, Giovanni and {Volonteri}, Marta and {Mannucci}, Filippo and {Maiolino}, Roberto and {Rodr{\'\i}guez Del Pino}, Bruno and {B{\"o}ker}, Torsten and {Bunker}, Andrew J. and {Charlot}, St{\'e}phane and {Willott}, Chris J. and {Carniani}, Stefano and {Curti}, Mirko and {Jones}, Gareth C. and {Kumari}, Nimisha and {Marshall}, Madeline A. and {Venturi}, Giacomo and {Saxena}, Aayush and {Scholtz}, Jan and {Witstok}, Joris},
        title = "{GA-NIFS: High number of dual active galactic nuclei at z {\ensuremath{\sim}} 3}",
      journal = {\aap},
         year = 2025,
        month = apr,
       volume = {696},
          eid = {A59},
        pages = {A59},
          doi = {10.1051/0004-6361/202453430},
archivePrefix = {arXiv},
       eprint = {2310.03067},
 primaryClass = {astro-ph.GA},
       adsurl = {https://ui.adsabs.harvard.edu/abs/2025A&A...696A..59P}
}

\end{document}